\documentclass[
 aip,
 amsmath,amssymb,
 reprint
]{revtex4-1}

\usepackage{graphicx}
\usepackage{dcolumn}
\usepackage{bm}

\usepackage[utf8]{inputenc}
\usepackage[T1]{fontenc}
\usepackage{mathptmx}
\usepackage{etoolbox}

\setcitestyle{super}
\usepackage[font=footnotesize,labelfont=bf,justification=RaggedRight]{caption}
\usepackage{float}
\usepackage{amsmath}
\usepackage{amssymb}
\usepackage{graphicx}
\usepackage{esint}
\usepackage{color}
\usepackage{comment}
\makeatletter    
\graphicspath{ {/} }            
\usepackage{amssymb}
\usepackage{hyperref}

\begin{document}

\preprint{AIP/123-QED}

\title[Gyrokinetic Simulations Compared with Magnetic Fluctuations Diagnosed with a Faraday-Effect Radial Interferometer-Polarimeter]{Gyrokinetic Simulations Compared with Magnetic Fluctuations Diagnosed with a Faraday-Effect Radial Interferometer-Polarimeter in the DIII-D pedestal}

\author{M. T. Curie}
 \affiliation{Institute for Fusion Studies, University of Texas at Austin, Austin, TX, 	78705}
  \affiliation{General Atomics, San Diego, CA, 85608}
  \affiliation{Princeton Plasma Physics Laboratory, Princeton, NJ 08540}
 \email{xingtian@fusion.gat.com}
 
  \author{D. R. Hatch}
 \affiliation{Institute for Fusion Studies, University of Texas at Austin, Austin, TX, 	78705}
 
  \author{M. Halfmoon}
 \affiliation{Institute for Fusion Studies, University of Texas at Austin, Austin, TX, 	78705}
 
   \author{J. Chen}
 \affiliation{Department of Physics and Astronomy, University of California, Los Angeles, Los Angeles, CA, 90095}
 
    \author{D.L. Brower}
 \affiliation{Department of Physics and Astronomy, University of California, Los Angeles, Los Angeles, CA, 90095}
 
  \author{E. Hassan}
 \affiliation{Oak Ridge National Laboratory, Oak Ridge, TN, 37830}
 \affiliation{Physics, Faculty of Science, Ain Shams University, EG, 11566}
 
   \author{M. Kotschenreuther}
 \affiliation{Institute for Fusion Studies, University of Texas at Austin, Austin, TX, 	78705}
 
   \author{S. M. Mahajan}
 \affiliation{Institute for Fusion Studies, University of Texas at Austin, Austin, TX, 	78705}
 
   \author{R. J. Groebner}
 \affiliation{General Atomics, San Diego, CA, 85608}

  \author{DIII-D team}
 \affiliation{General Atomics, San Diego, CA, 85608}
 \affiliation{See the author list of 'DIII-D research advancing the physics basis for optimizing the tokamak approach to fusion energy’ by M. Fenstermacher et al. to be published in Nuclear Fusion (Accepted on 14 October 2021)}

\date{\today}

\begin{abstract}
Experimental data on electromagnetic fluctuations in DIII-D, made available by the Faraday-effect Radial Interferometer-Polarimeter (RIP) diagnostic\cite{rip}, is examined in comparison with detailed gyrokinetic simulations using Gyrokinetic Electromagnetic Numerical Experiment (GENE). The diagnostic has the unique capability of making internal measurements of fluctuating magnetic fields $\frac{\int n_e \delta B_r dR}{\int n_e dR}$. Local linear simulations identify microtearing modes (MTMs) over a substantial range of toroidal mode numbers (peaking at $n=15$) with frequencies in good agreement with the experimental data. Local nonlinear simulations reinforce this result by producing a magnetic frequency spectrum in good agreement with that diagnosed by RIP.  Simulated heat fluxes are in the range of experimental expectations.  However, magnetic fluctuation amplitudes are substantially lower than the experimental expectations.  Possible sources of this discrepancy are discussed, notably the fact that the diagnostics are localized at the mid-plane---the poloidal location where the simulations predict the fluctuation amplitudes to be smallest.  Despite some discrepancies, several connections between simulations and experiments, combined with general criteria discriminating between potential pedestal instabilities, strongly point to MTMs as the source of the observed magnetic fluctuations.
\end{abstract}

\maketitle

\section{Introduction}

This paper presents comparisons between gyrokinetic simulations and novel measurements of internal magnetic fluctuations in the pedestal of a DIII-D discharge.

The pedestal forms when virulent fluctuations are suppressed at the edge of the Tokamak.  Due to the large impact of the pedestal on confinement, the origin and role of the residual fluctuations is a key question for developing a comprehensive understanding of the pedestal and projecting its behavior to foreign parameter regimes like those envisioned for burning plasmas.  

Gyrokinetic simulations have demonstrated ever greater fidelity in comparison with experimental observations in the core plasma.~\cite{Core,doi:10.1063/1.4919022,osti_1779308,doi:10.1063/1.2895408,doi:10.13182/FST15-182} 

Notably, multi-channel transport, fluctuation levels, cross phases, and turbulent spectra have all been favorably compared between simulation and experiment.  Although gyrokinetic pedestal simulations have been advancing \cite{Haskey_2022,PhysRevLett.108.135002,Hatch_2016,Hatch_2017,PhysRevLett.123.115001,doi:10.1063/1.5124986,Hatch_2021,osti_1776857,doi:10.1063/5.0037246,curie_thesis}
, detailed comparisons with fluctuation amplitudes are rare.  This paper reports on some of the first such comparisons in the pedestal.  Such a comparison is enabled by novel measurements of internal magnetic fluctuations from the recently-developed Faraday-effect Radial Interferometer-Polarimeter (RIP)\cite{rip}.  This diagnostic directly measures the radial magnetic fluctuation levels across the mid-plane.  In contrast, earlier fluctuation data from magnetic pick-up coils (external to the plasma) were capable of determining fluctuation frequencies but not fluctuation amplitudes.  

In this paper, we describe gyrokinetic simulations \cite{Jenko_GENE,GORLER20117053}
of a DIII-D pedestal.  The simulations identify microtearing modes (MTMs) as the main ion-scale instability.  Nonlinear simulations produce frequency spectra, transport levels, and collisionality dependence in good agreement with these experimental measurements.  However, the simulations predict magnetic fluctuation levels lower than the experimental expectation.  Possible sources for this discrepancy are discussed.  We also consider the ratio of magnetic fluctuations to density fluctuations.  Although there is also some discrepancy between experiment and simulation for these measurements, the fluctuation's features are generally consistent with MTM and inconsistent with other instabilities.  
    
The structure of the paper is outlined as follows: 

\begin{itemize}
    \item  \textbf{Background} of the theory and experiment. The significance of MTM will be discussed. And the brief description of the Faraday-effect Radial Interferometer-Polarimeter (RIP) will be provided.
    
    \item \textbf{Linear Investigation} We study the linear eigenmode spectrum, identifying the salient modes and making preliminary comparisons with the RIP frequency spectrum.
    
    \item \textbf{Nonlinear Investigation} A set of direct comparisons between experiment and local nonlinear MTM simulations will be performed including scans of the frequency spectrum and collisional dependence. Local nonlinear ETG (electron temperature gradient) simulations are conducted as a contrast with local nonlinear MTM simulations to study spatial dependence and fluctuation ratio. A comparison of electron heat transport between the experiments and simulations is performed by adding the electron heat transport of ETG and MTM from simulations.
\end{itemize}

\section{Theory and Experimental Background}
\label{background}

Recent work has elucidated the potential roles of various transport mechanisms in the pedestal.  Historically, Kinetic Ballooning Mode (KBM) has been proposed as the salient transport mechanism in the pedestal due to its utility in the EPED model and simulation results suggesting that pedestal profiles lie near the KBM limit~\cite{Saarelma_2017,Hatch_2015,Hatch_2016,Canik_2013}.
Recent work, however, has determined that a variety of mechanisms is required to account for pedestal transport in all channels \cite{Kotschenreuther_2019,TPT19,Joel_prl,Fenstermacher_2022}.  In particular, edge modeling in simulations \cite{TPT19} typically predicts that the effective electron heat diffusivity far exceeds the electron particle diffusivity: $D_e / \chi_e \ll 1$.  In contrast Magnetohydrodynamic (MHD) modes, like KBM, are characterized by $D_e \sim \chi_e$.  Moreover, KBM has been found to be in a second stability regime in some scenarios.  While these observations do not eliminate KBM as a prospective transport mechanism, it does demand investigation of additional transport mechanisms and suggests that other mechanisms are necessary to describe pedestal transport in all transport channels.  Table~\ref{ch:finger} shows the distinctive `fingerprints' of various prospective transport mechanisms, notably MTM, which is the main focus of this paper.  There is growing evidence that in the electron heat channel, electron temperature gradient (ETG) turbulence and MTMs fill this role \cite{Hatch_2016,Ehab_MTM,Curie_2022_SLiM,Kotschenreuther_2019,RIP_Chen_POP_2020}, likely often in tandem \cite{Hatch_2021,Pueschel_2020}.  

\begin{table}[H]
    \centering
    \begin{tabular}{|c|c|c|c|c|c|c|c|}
         \hline
          Mode Type & $\chi_i/\chi_e$ & $D_e/\chi_e$ & $D_z/\chi_e$ & $Q_{em}/Q_{es}$ & Shear-\\&&&&&suppressed? \\
         \hline
          MTM & $\sim 0$ & $\sim0$ & $\sim0$ & $>1$ & Sometimes \\
         \hline
          ETG & $\sim0$ & $\sim0$ & $\sim0$ & $<1$ & No \\
         \hline
          MHD-like & $\sim1$ & $\sim2/3$ & $\sim2/3$ & $>1$ & No\\
         \hline
          ITG/TEM & $\geq 1$ & $-0.2-1$ & $\sim1$ & $<1$ & Usually\\
         \hline
\end{tabular}
\caption[font=5]{Theoretical estimates of transport ratio for different instabilities, $D_s$ is particle diffusion coefficient for species 's', $\chi_s$ is heat diffusion coefficient for species 's', $Q_{em}$ and $Q_{em}$ are electromagnetic and electrostatic heat flux respectively. }
\label{ch:finger}
\end{table}

High-frequency magnetic fluctuations originating in the pedestal have long been observed with no clear explanation for the underlying physical mechanism
\cite{Ahmed_nonlinear, Dominski_nonlinear2, Dominski_nonlinear2,Perez_2003,PhysRevLett.112.115001,doi:10.1063/1.4921148,Laggner_2016,LAGGNER2019479,doi:10.1063/5.0040306,rip,RIP_Chen_POP_2021,MTM_power}.  Recently, gyrokinetic simulations have, with increasing certainty, established MTMs as the underlying mechanism for at least a major class of these fluctuations.  Ref.\cite{Hatch_2016} demonstrated that gyrokinetic simulations of MTMs can produce experimentally relevant transport levels for a JET discharge, and Refs.\cite{Kotschenreuther_2019,Hatch_2021,Ehab_MTM} demonstrated close agreement with the distinctive quasi-coherent frequency bands measured by magnetic pick-up coils.  Refs. \cite{Hatch_2021,Ehab_MTM} provided an explanation for the distinctive band structure: MTMs arise at the peak of the electron diamagnetic frequency $\omega_{*,e}= k_y \rho_s c_s (1/L_{Te}+1/L_{ne})$ profile only for mode numbers whose rational surfaces align with this peak.  This numerical observation has been further elucidated by basic theory, which demonstrates that this behavior is due to the profile variation in the pedestal \cite{Joel_prl,Curie_2022_SLiM}.  




\subsection{Internal Measurements of Magnetic Fluctuations on DIII-D}

Recent experimental work based on the RIP diagnostic has further established MTM as the mechanism for magnetic fluctuations by demonstrating several parameter dependencies that are consistent with MTM and inconsistent with other candidates \cite{RIP_Chen_POP_2021}.  

\begin{figure}[ht]
        \includegraphics[width=0.3\textwidth]{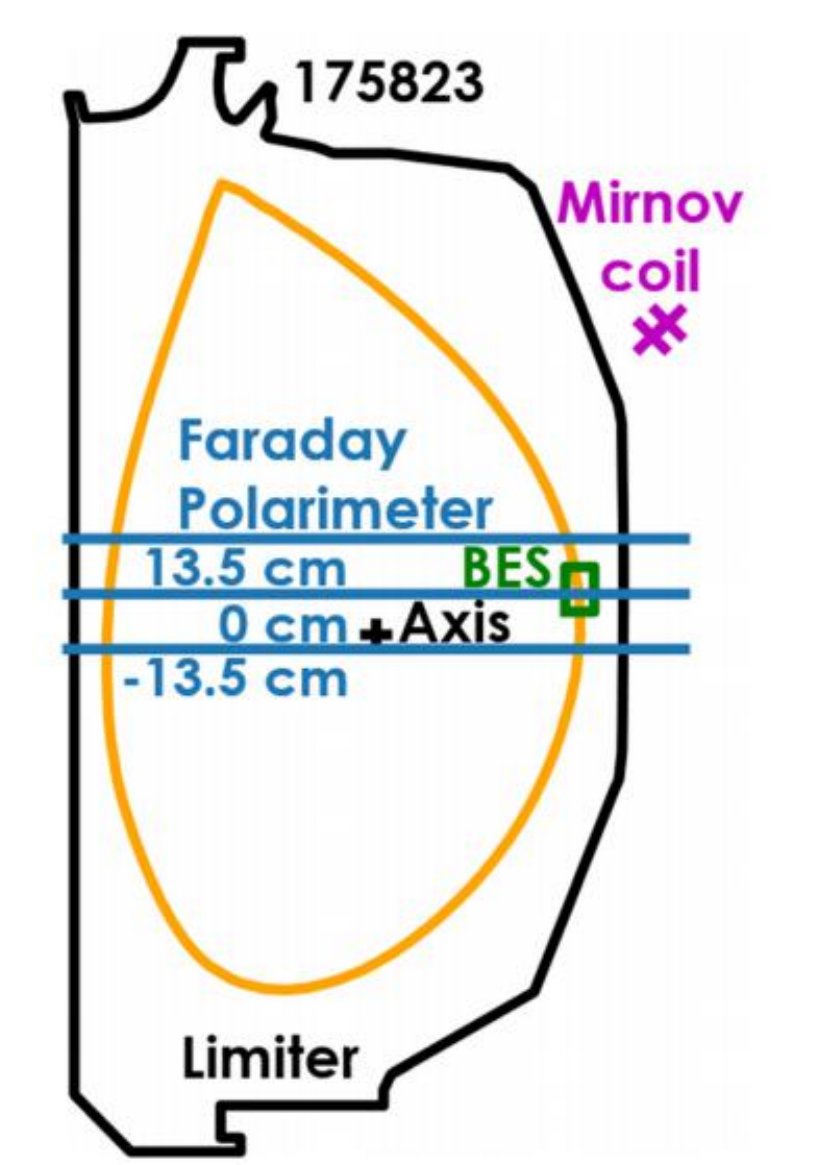}
        \centering
        \caption[font=5]{This figure shows the schematic of relevant diagnostic: 
        3 horizontal lines show 3 horizontal chords of RIP diagnostic at the mid-plane (0 cm), 13.5 cm above the mid-plane, and 13.5 cm below the mid-plane. 
        The green box is Beam Emission Spectroscopy (BES) located at the pedestal. 
        This figure is reused from Chen (2021) \cite{RIP_Chen_POP_2021}}
        \label{fig:RIP}
\end{figure}

Radial-interferometer-polarimeter (RIP) is a newly developed Faraday-effect polarimeter in DIII-D, which measures the electron density-weighted magnetic fluctuation along horizontal chords at and around the mid-plane as the Fig.~\ref{fig:RIP} shows. The measured line-averaged radial magnetic fluctuation amplitude from the RIP can be expressed as 

\begin{equation}
\delta \bar{B}_{r} \approx \delta \bar{B}_{R} \equiv \frac{\sqrt{\overline{\left(\int n_{e} \delta B_{R} d R\right)^{2}}}}{\int n_{e} d R}
\label{eq:RIP_Br}
\end{equation}



\noindent where $\sqrt{\overline{\left(\int n_{e} \delta B_{R} d R\right)^{2}}}$ is the root mean square (RMS) of the electron density-weighted internal magnetic fluctuation. \cite{RIP_Chen_POP_2020} $n_e$ is electron density, $\delta B_R$ is the magnetic fluctuation along horizontal chords. $dR$ is differential element of radius. 

The detailed experiment setup can be found in Chen (2021)\cite{RIP_Chen_POP_2021}

This diagnostic provides a direct measurement of the absolute magnetic fluctuation amplitude inside the plasma. Although this diagnostic is not localized radially, magnetic fluctuations in the core generally have lower frequencies ($\sim1kHz$) so that frequency filtering (in the range $150\sim 500 kHz$) can effectively extract the relevant pedestal fluctuations. And the fact that the RIP observed frequency band responds to ELMs (edge-localized mode) and correlates to local measurements (BES) of electron density at the pedestal further proves that the $150\sim 500 kHz$ frequency band is localized at the pedestal.\cite{RIP_Chen_POP_2021} 

For more experimental aspects of this research, the reader is referred to the following publications: \cite{rip,RIP_Chen_POP_2021,RIP_Chen_POP_2020} 

\subsection{DIII-D Discharge 175823}

\begin{figure}[h]
\centering
        \includegraphics[width=0.45\textwidth]{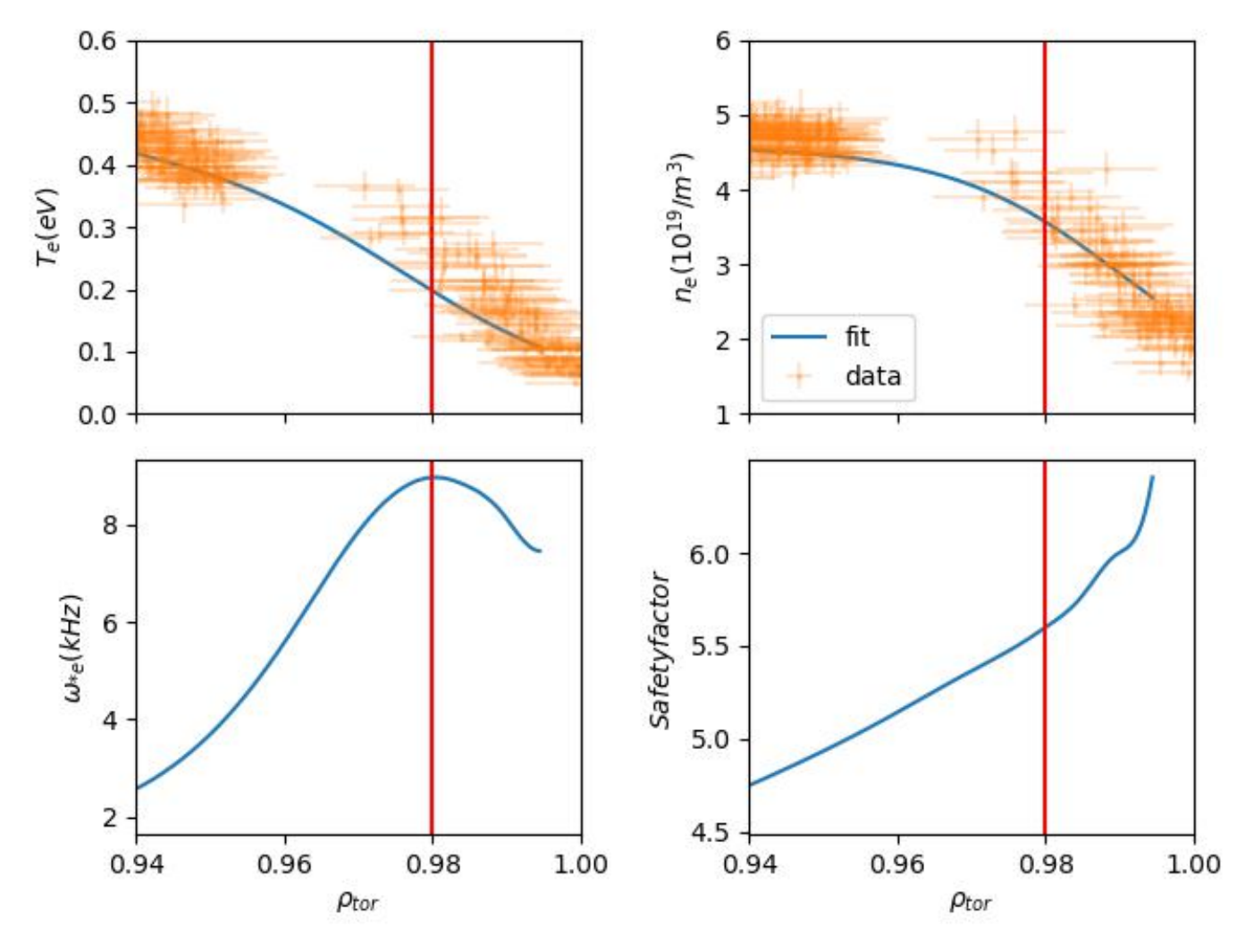}
        \caption[font=5]{The experimental equilibrium profile: electron temperature $T_e$ (top left), electron diamagnetic frequency $\omega_{*e}$ (bottom left), electron density $n_e$ (top right), safety factor $q$ (bottom right). The orange data points are experimental data, and the blue curves are the fitted profile based on the experimental data (orange dots). And the red vertical lines mark $rho_{tor}=0.98$, where the local simulations are conducted in the article. }
        \label{fig:EFIT}
\end{figure}

All the simulations conducted in the article are based on DIII-D H-mode shot 175823. This discharge exhibits type-I ELMs with ELM frequency $\sim 100 Hz$. The plasma is heated via neutral beam injection. Its magnetic geometry has an upper single-null. Its line averaged density is $5.5\times10^{19}/m^3$ and $H_{98}\sim 1.1$. The further detail of this discharge can be found in Chen (2021) \cite{RIP_Chen_POP_2021}. 
The profiles of density, temperature, electron diamagnetic frequency, and safety factor $q$ are shown in Fig.~\ref{fig:EFIT}, where $\omega_{*e}$ is the electron diamagnetic frequency, $q$ is the safety factor $q\equiv\frac{d\Psi_{T}}{d\Psi_{P}}$, $\hat{s}$ is the magnetic shear $\hat{s}\equiv \frac{r}{q}\frac{dq}{dr}$, $L_{T_e}$ is the length scale of the electron temperature gradient, $L_{n_e}$ is the length scale of the electron density gradient. $a$ is the minor radius, $r$ is the minor radial length for the given location, $\rho_s$ is the ion Larmor radius, $c_s$ is the speed of sound in plasma. 


The RIP diagnostic provides line integrated radial magnetic fluctuation amplitudes~\cite{rip} $\frac{\int n_e \delta B dR}{\int n_e dR}$ across the mid-plane.  This provides us with a unique insight into the internal measure of magnetic fluctuations. 



The magnetic frequency spectrum is one of the key experimental signatures that we wish to compare with simulation results.  Consequently, we need an estimate of the Doppler shift of the fluctuations.  This is provided by CXRS (Charge eXchange Recombination Spectroscopy), which determines the radial electric field from the radial force balance of the carbon species.  The resulting Doppler shift for instability with toroidal mode number $n$ ~\cite{Hatch_2021}:
\begin{equation}
    \Omega_{Doppler} = \frac{n E_r}{R B_\theta} 
\end{equation}
where $E_r$ is the radial electric field, $B_\theta$ is the poloidal magnetic field, $R$ is the major radius, n is the toroidal mode number.  For this discharge, taking $n=1$, the pedestal rotation frequency is calculated to be $E_r / RB_\theta \sim 10 krad/s$.  There are significant error bars on this estimate, which are difficult to quantify.  We take a rough estimate of the error bars by examining uncertainties for the pressure gradient, which makes the main contribution to $E_r$ in the radial force balance equation.  This exercise suggests that the contribution of the pressure gradient to the rotation is $\sim 25 \pm 10 krad/s$.  Consequently, we test two values for the rotation in this work: $\Omega_{Doppler} = 10 krad / s$ and $\Omega_{Doppler} = 20 krad / s$.

\section{Gyrokinetic Simulations}
\label{simulations}

The profile and equilibrium for DIII-D discharge 175823 is constructed by kinetic EFIT\cite{kinetic_EFIT}. 
We use the reconstructed profiles and equilibrium for DIII-D discharge 175823 for all simulations.  In this paper, we focus on local flux tube simulations.  Previous work has shown that global simulations are required to capture the sensitive selection of frequency bands and toroidal mode numbers that are often observed in magnetic fluctuation data~\cite{Hatch_2021,Joel_prl}.  However, this discharge has rather high magnetic shear ($\hat{s}=5$), and the fluctuations are characterized by single broadband representing multiple toroidal mode numbers ranging from 10 to 20.  Consequently, a local approximation is reasonable (with caveats for nonlinear simulations, which will be discussed below).  Previous results indicate that the global modes localize at the peak of the $\omega_{*e}$ profile, providing a clear radial location at which to perform the local analysis.  

Numerical details as well as a discussion of convergence tests can be found in the Appendix.

\subsection{Linear Simulations}
\label{nonlinear}

We first describe investigations into the linear spectrum of unstable modes in the pedestal of DIII-D discharge 175823.  As will be seen, this is already sufficient to make contact with experimental observations.

\begin{figure}[ht]
        \includegraphics[width=0.45\textwidth]{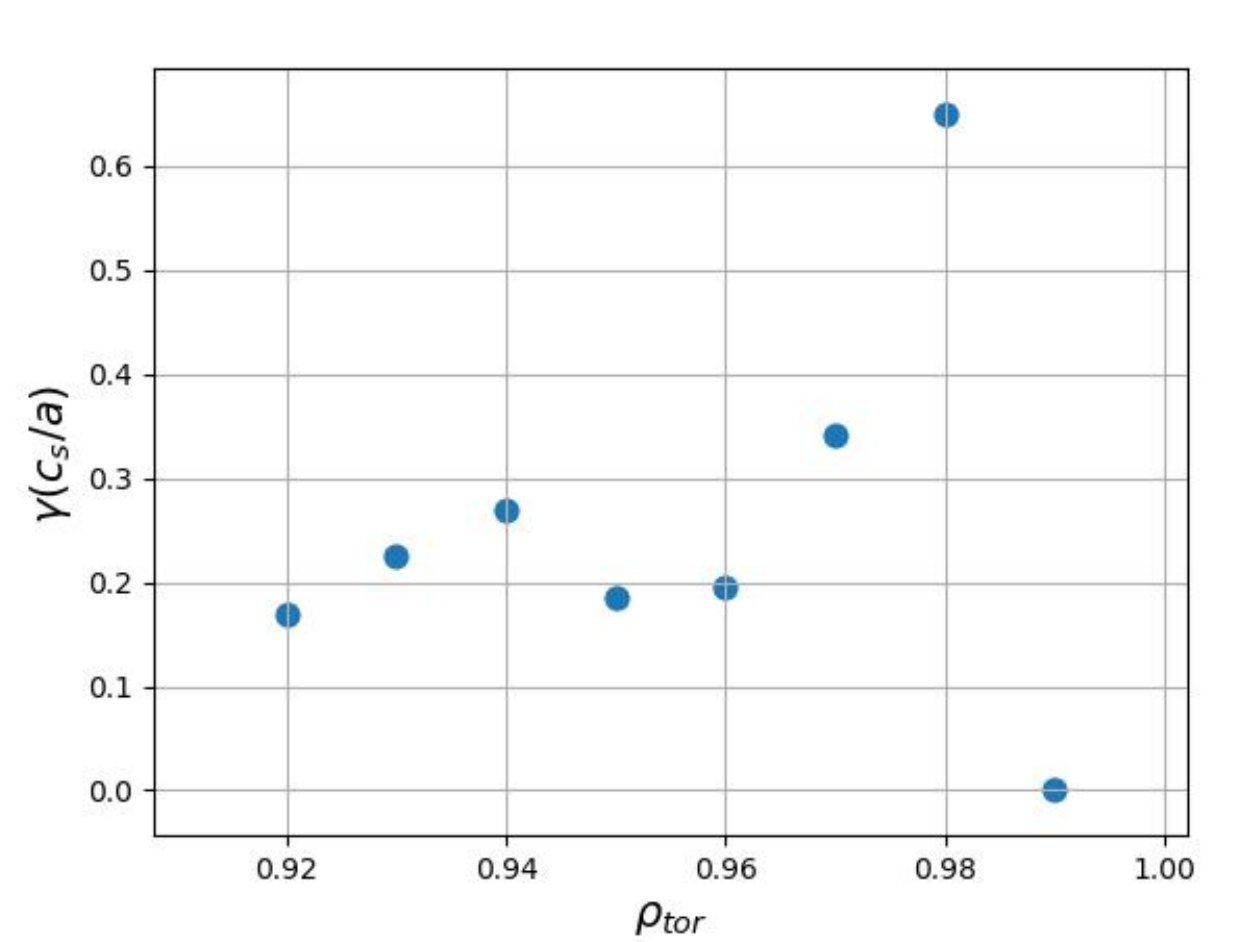}
        \centering
        \caption[font=5]{Local linear simulation results: Growth rate of unstable MTM over radial location. }
        \label{fig:gamma}
\end{figure}

Fig.~\ref{fig:gamma} shows local linear simulations scanning the radial domain of the pedestal (from $\rho_{tor}=0.92\sim 0.99$), showing the growth rate of the most unstable MTM at each location. As expected, the instability is most unstable at the mid-pedestal $\rho_{tor}=0.98$, where $\omega_{*e} = k_y \rho_s c_s (1/L_{T_e}+1/L_{n_e})$ peaks. Refs.~\cite{Hatch_2021, Joel_prl, Ehab_MTM} demonstrate that (1) pedestal MTMs peak in the potential well created by the $\omega_*$ profile, and (2) at low toroidal mode number and/or low magnetic shear, MTM stability is extremely sensitive to the alignment of this potential well with mode rational surfaces.  For this scenario, the combination of high magnetic shear ($\hat{s} \sim 5$ and relatively high toroidal mode numbers ($n>10$) produce MTMs that are not sensitive to these considerations.  Consequently, consistent with the results in Fig.~\ref{fig:gamma} we proceed with a detailed local analysis at a single location -- $\rho_{tor}=0.98$. Relevant parameters at mid-pedestal are shown in Table \ref{ch:basic_para} 

\begin{table}[H]
    \centering
    \begin{tabular}{|c|c|c|c|c|c|c|c|}
         \hline
         Pulse & $q$ &  $T_e$ (keV) & $n_e (10^{19}/m^3)$ & $\nu_{ei} (c_s/a)$\\
         \hline
         175823 & 5.59 & 0.197 & 3.57 & 7.38  \\
         \hline \hline
         $\rho_{tor}$ & $\hat{s}$& $a/L_{T_{e}}$ & $a/L_{n_{e}}$ & $\beta(10^{-4})$\\
         \hline
         0.98 & 4.94 & 37.0 & 17.1 & 7.24\\
         \hline \hline
         R (m) & $a(m)$& $B_0$ (T) & $\rho_s$ (mm) & $c_s$ (km/s) \\
         \hline
         1.73 & 0.739 & 1.977 &1.023 & 69.1\\
         \hline
    \end{tabular}
    
    
    \begin{tabular}{|c|c|c|c|c|}
         \hline
         $\Omega_g$ ($c_s$/a)\\
         \hline
         9362\\
         \hline
    \end{tabular}
\caption[font=5]{Basic parameters at the mid-pedestal: where q is the safety factor $q\equiv\frac{d\Psi_{T}}{d\Psi_{P}}$, $\nu_{ei}$ is the electron-ion collision frequency, $\beta$ is ratio of the plasma pressure to the magnetic pressure, $\hat{s}$ is the magnetic shear, $L_{T_e}$ is the length scale of the electron temperature gradient, $L_{n_e}$ is the length scale of the electron density gradient, $R$ is the major radius, $a$ is the minor radius, $\rho_s$ is the ion Lamar radius, $c_s$ is the speed of sound in plasma, $\Omega_g$ is the ion gyro-frequency, $B_0$ is the magnetic field strength}
\label{ch:basic_para}
\end{table}

\begin{figure}[h]
        \includegraphics[width=0.45\textwidth]{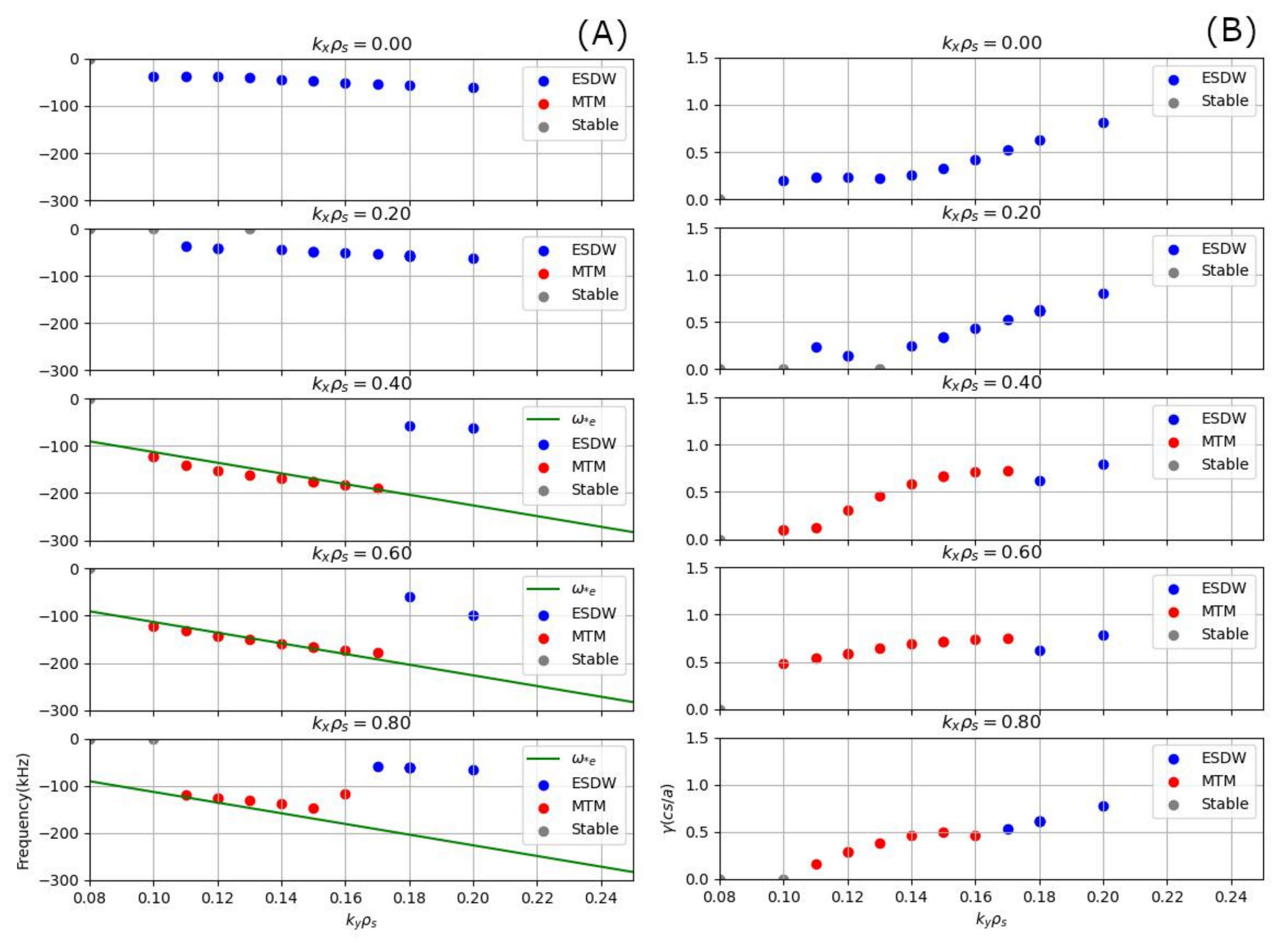}
        \centering
        \caption[font=5]{Local linear simulation results: plot (A) on the left hand side shows the growth rate over mode number ($k_y\rho_s=0.00776\times n$, where $n$ is toroidal mode number). Red dots are the MTMs, blue dots are ESDWs, mode numbers with no unstable modes are marked with grey dots. 
        plot (B) on the right hand side shows frequency in plasma frame over mode number. Red dots are the MTMs, blue dots are ESDWs, mode numbers with no unstable modes are marked with grey dots. The green lines are the electron diamagnetic frequency $\omega_{*e}$. The negative frequency represents the frequency that is in the electron direction}
        \label{fig:LL_f_gamma}
\end{figure}

Fig.~\ref{fig:LL_f_gamma} (A) shows growth rates scanned versus $k_y \rho_s$ and $k_x \rho_s$ at $\rho_{tor}=0.98$ (the translation from $k_x$ to ballooning angle $\theta_0$ is as follows: $\theta_0=\frac{k_x}{\hat{s}k_y}$). The definition of $k_y$ is $k_y=\frac{n_{tor} q}{a}$, where $n_{tor}$ is toroidal mode number. As in earlier work~\cite{Hatch_2016, Ehab_MTM}, we find that the modes peak at finite ballooning angle $\theta_0$ (finite $k_x$).  The MTMs coexist with an electrostatic drift wave (ESDW, denoted with blue symbols) with strong electron signatures---predominantly electron heat flux and frequencies in the electron direction.  The main distinction between ESDW and MTM is the very weak component of electromagnetic heat flux in the former.

The MTM is subdominant at $\theta_0 = 0$ and becomes dominant only as ballooning angle increases.  Fig.~\ref{fig:LL_f_gamma} (B) shows frequencies in the plasma frame in $kHz$.  
Although these local linear simulations are in the plasma frame, one can already note the high frequencies ($>100kHz$) characteristic of the MTMs.

\begin{figure}[ht]
        \includegraphics[width=0.45\textwidth]{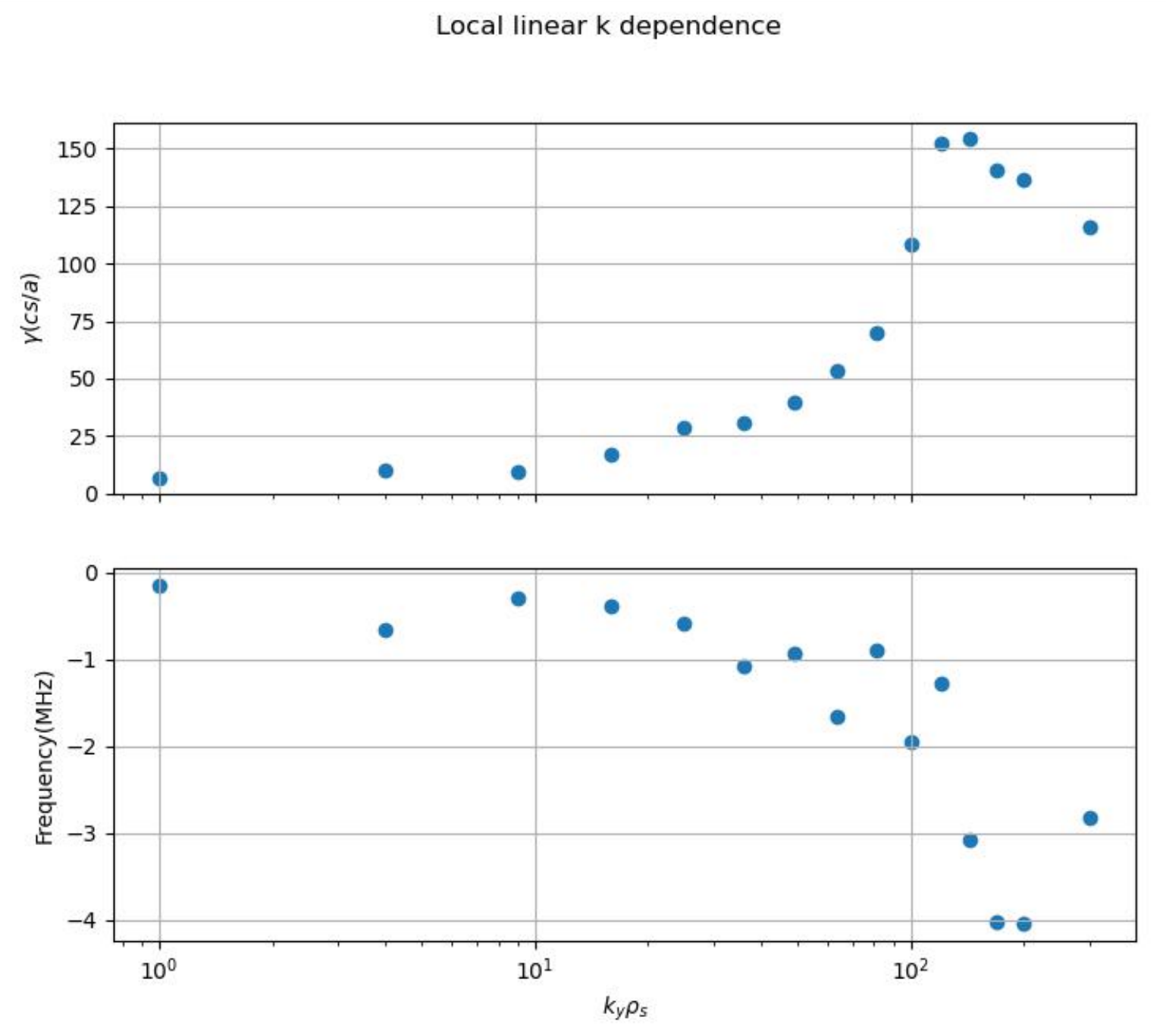}
        \centering
        \caption[font=5]{Local linear simulation results for high $k_y$ electron scales: frequency in plasma frame and growth rate in electron scale. All instabilities shown in these plots are ETG}
        \label{fig:LL_ETG}
\end{figure}

For reference, the linear spectrum for high $k_y$ electron scales is shown in Fig.~\ref{fig:LL_ETG}.  The corresponding single-scale nonlinear ETG simulations (described briefly below) produce low transport levels, suggesting that the focus on MTM is well-justified.  

\begin{figure}[ht]
        \includegraphics[width=0.45\textwidth]{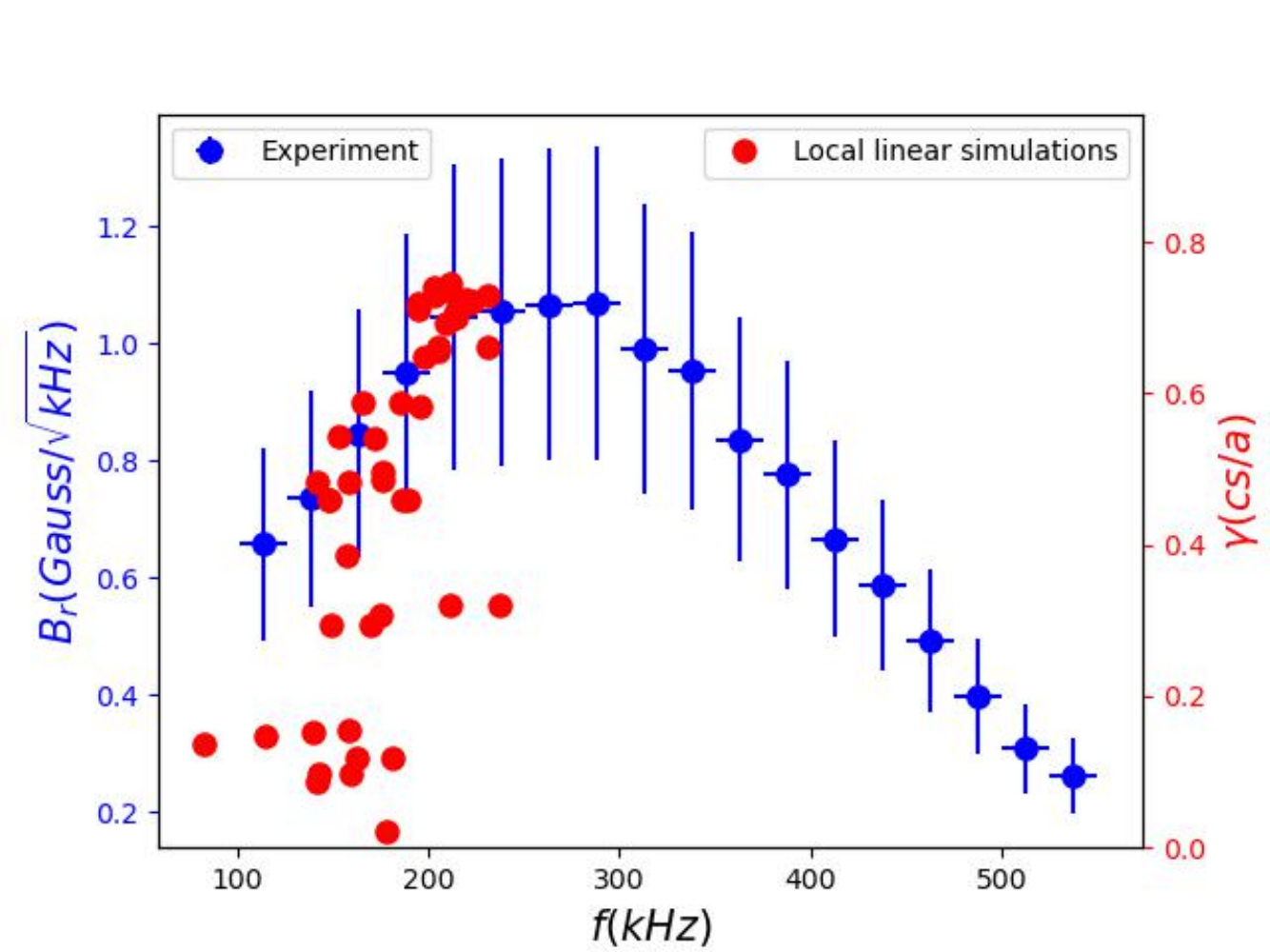}
        \centering
        \caption[font=5]{Comparison of local linear simulation of MTM with experimental observation: the red dots are lab frame frequency VS growth rate of the local linear simulations, the blue dots are the experimental observation of magnetic fluctuation amplitude}
        \label{fig:LL_exp}
\end{figure}

As the first point of experimental comparison, we show the frequency spectrum as calculated from the RIP diagnostic in comparison with the frequencies of the linear MTMs in Fig.~\ref{fig:LL_exp}.  In this figure, the experimental frequency spectrum represents the magnetic fluctuation amplitude, while the linear growth rates are used for the linear modes.  The frequencies are translated into the lab frame using $\frac{E_r}{R B_{pol}} = 1.5 kHz$, as described above. 
The frequencies that correspond to the peak growth rates are in good agreement with the peak of the experimental spectrum.  Note from Fig.~\ref{fig:LL_f_gamma} (A) that, as $k_y$ increases, the electrostatic modes dominate just beyond the peak in MTM growth rates (growth rates are shown in Fig.~\ref{fig:LL_f_gamma} [B] ).  This suggests that higher frequency MTMs corresponding to the upper tail of the experimental spectrum are likely still unstable but subdominant to the electrostatic modes.  

\begin{figure}[ht]
        \includegraphics[width=0.45\textwidth]{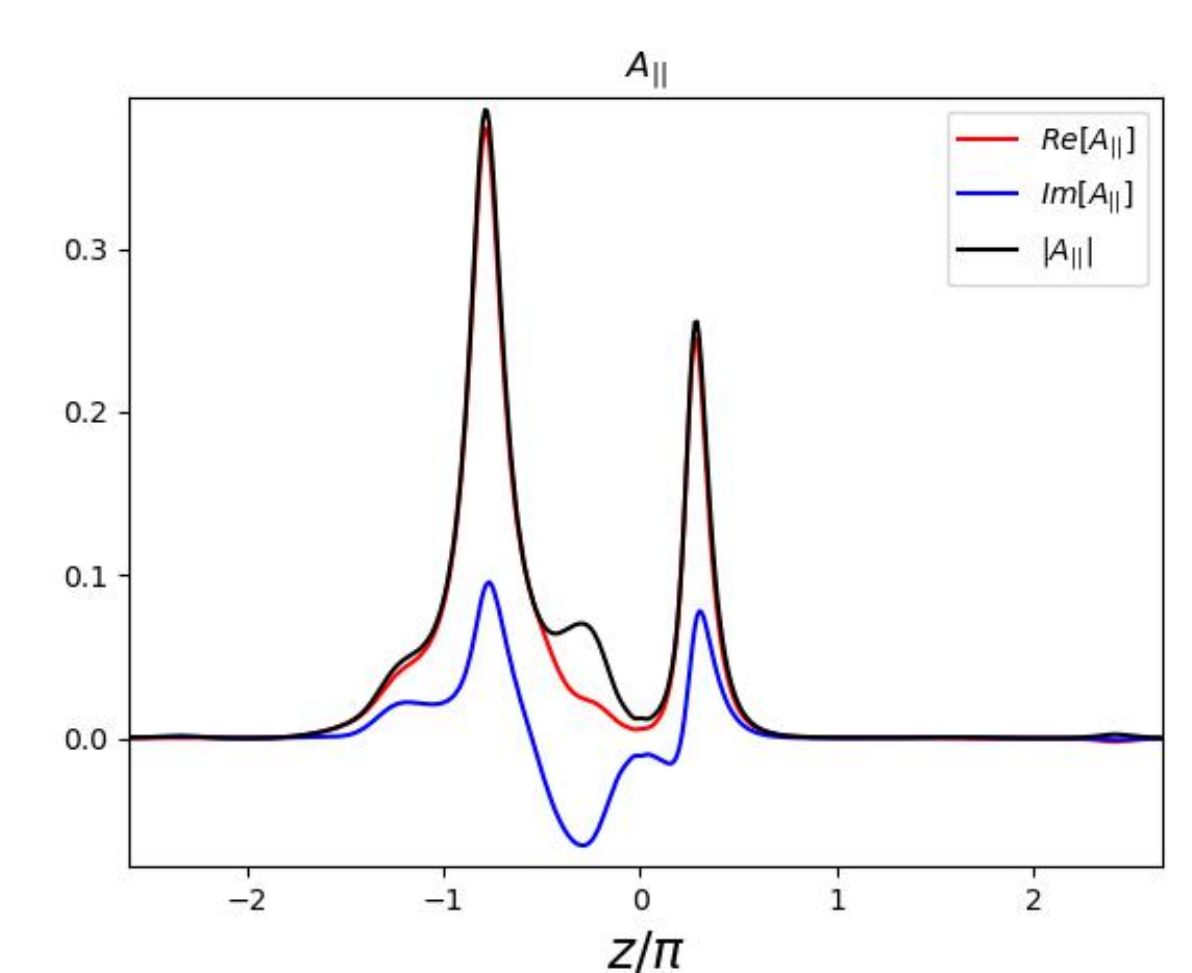}
        \centering
        \caption[font=5]{$A_{\|}$ Mode structure of MTM with $k_x\rho_s=0.6$, $k_y\rho_s=0.14$. The y axis has arbitrary unit that is proportional to $A_{\|}$}
        \label{fig:MTM_Apar_mode}
\end{figure}

\begin{figure}[ht]
        \includegraphics[width=0.45\textwidth]{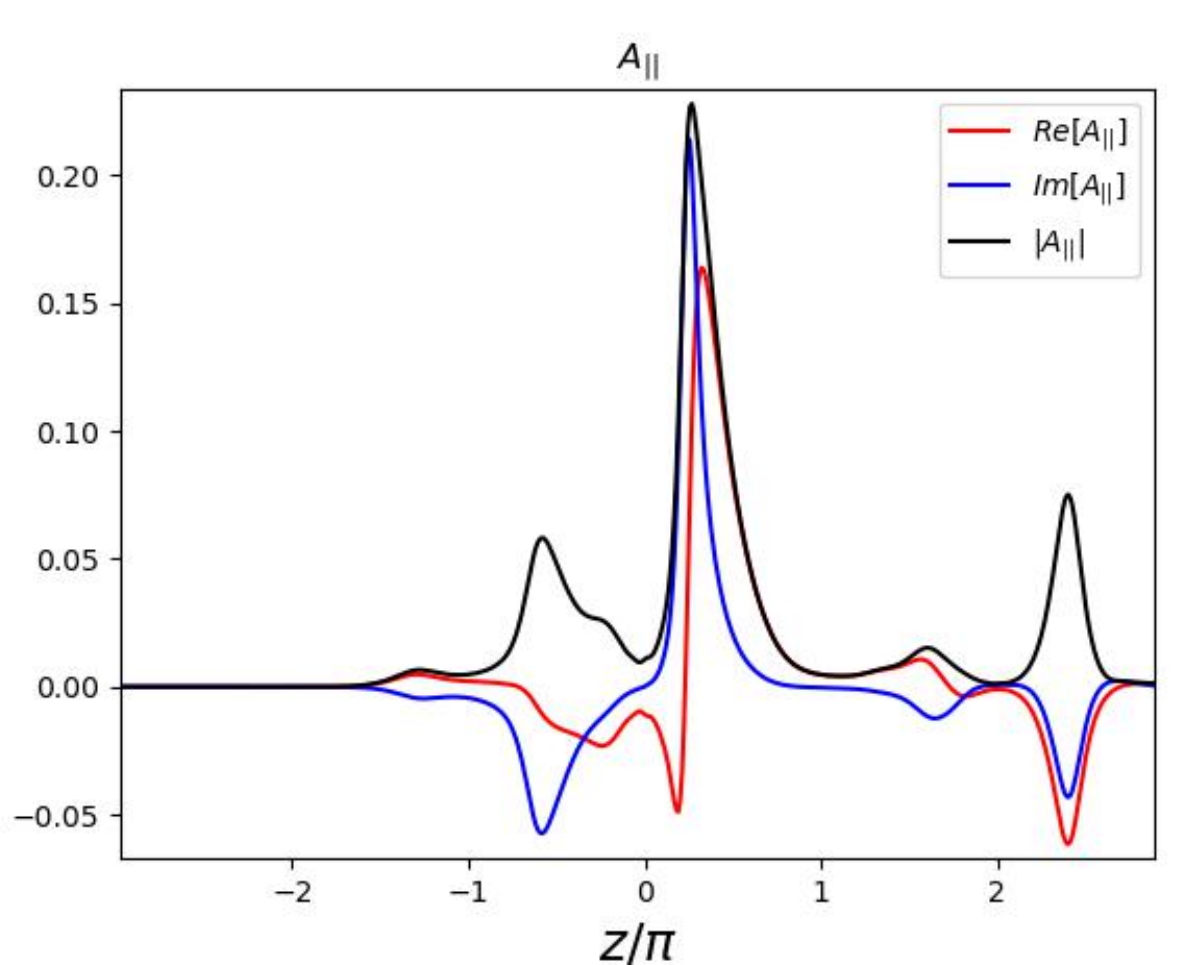}
        \centering
        \caption[font=5]{$A_{\|}$ Mode structure of ESDW with $k_x\rho_s=0.6$, $k_y\rho_s=0.2$. The y axis has arbitrary unit that is proportional to $A_{\|}$}
        \label{fig:ETG_Apar_mode}
\end{figure}

Figure \ref{fig:MTM_Apar_mode} and  Figure \ref{fig:ETG_Apar_mode} show the $A_{\|}$ mode structures of representative MTM and ESDW modes, respectively, from the local linear simulations.  The mode structure of MTM and ESDW are distinct from one another: the MTM has (roughly) an even parity of $A_{\|}$ (imaginary part or real part) around ($\theta = 0$), while ESDW has (roughly) odd parity of $A_{\|}$ (imaginary part or real part) \cite{MJ_ESDW}. Aside from that, the modes are not the simple structures at the outboard mid-plane familiar from core-like scenarios.  For example, they extend further in $\theta$ and generally exhibit more structure.  


\begin{figure}[ht]
        \includegraphics[width=0.45\textwidth]{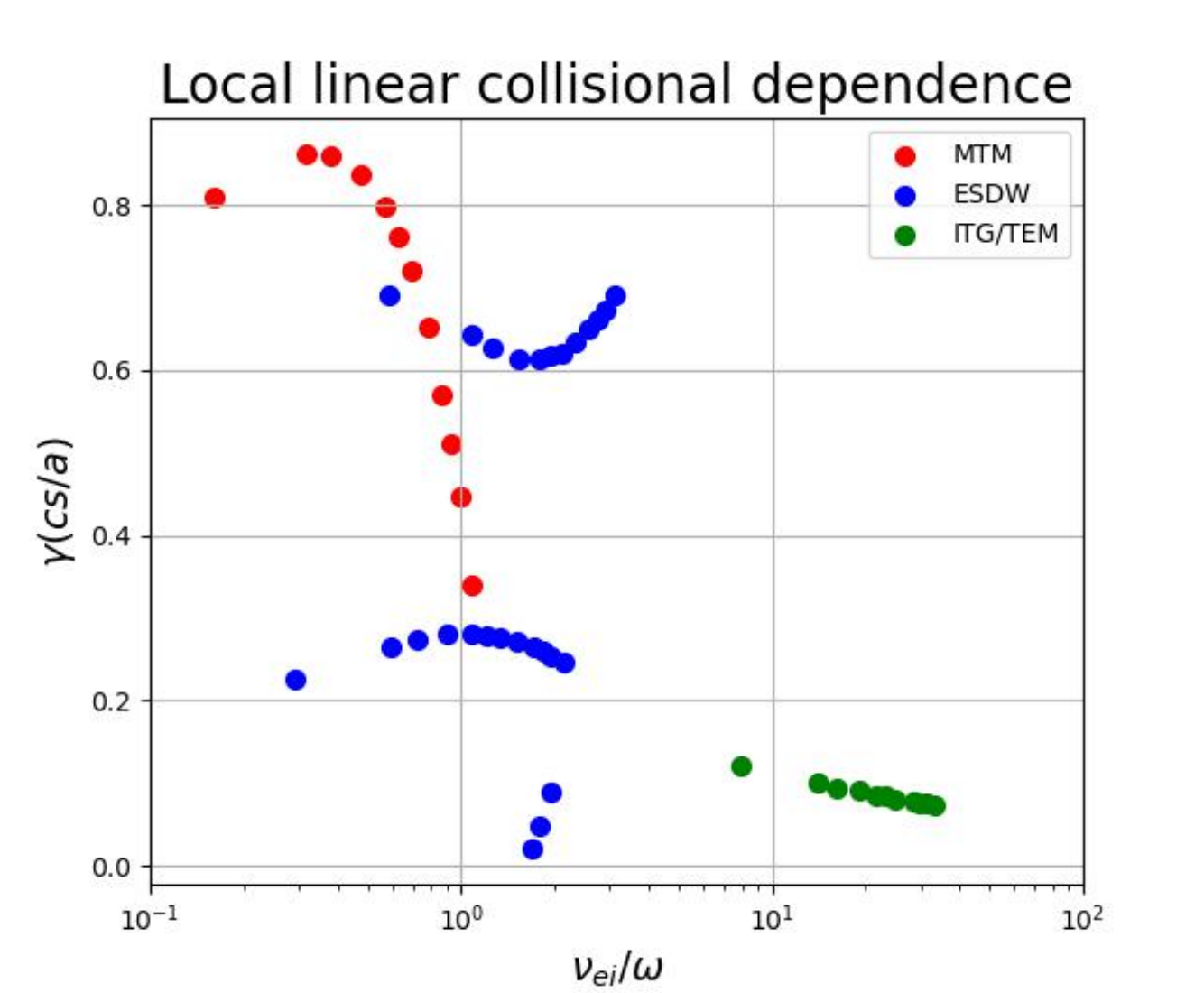}
        \centering
        \caption[font=5]{Local linear collisionality scan of at the $\rho_{tor}=0.98$, $k_y\rho_s=0.14$, $k_x\rho_s=0.6$. $\nu_{ei}$ is the electron-ion collision frequency, $\omega$ is the mode frequency}
        \label{fig:EV_nu}
\end{figure}

As a final investigation into the linear properties of the mode, we investigate the collisionality dependence of the growth rates as shown in Fig.~\ref{fig:EV_nu}.  
The MTM growth rates are plotted in addition to those of the electrostatic mode using results from the GENE eigenvalue solver.  The MTMs exhibit the expected non-monotonic collisionality dependence~\cite{gladd,MTM_RH,Joel_prl}, while the other modes exhibit much milder collisionality dependence.  The simulations are conducted at $\rho_{tor}=0.98$, $k_y\rho_s=0.14$, $k_x\rho_s=0.6$ where the MTM growth rate peaks.  The experimental magnetic fluctuation levels also exhibit non-monotonic behavior with respect to collisionality.  We will return to this point in the next section in the context of comparing a nonlinear collisionality scan with experimental data.  


\subsection{Nonlinear Simulations}

A set of local nonlinear simulations has been conducted at the location where the growth rate of MTM maximized --- $\rho_{tor}=0.98$. Nonlinear simulations in the pedestal are very challenging due to strongly shaped flux surfaces, proximity to MHD limits, and extreme gradients.  As will be seen, these simulations do make some clear connections with experimental observations.  However, they also have certain limitations.  Most notably, following an intermediate time period with saturated heat fluxes, the simulations transition to a period of runaway growth toward unrealistic transport levels.  Despite these limitations, we have decided to present these simulations in order to (1) inform the community regarding numerical issues in this challenging parameter regime, and (2) analyze the intermediate saturation phase in comparison with experimental observations.  The long-time runaway is the consequence of low-$k_y$ modes that grow without bound.  It is our tentative conclusion that the intermediate saturation phase is at least qualitatively meaningful and that global effects are critical for stabilizing these low-n modes---i.e. global simulations would constrain the low-$k_y$ fluctuations to radial scales consistent with the width of the pedestal and the width of the driving range of the modes.  Global simulations exploring these effects are currently under investigation and will be published elsewhere.    

\begin{figure}[ht]
        \includegraphics[width=0.45\textwidth]{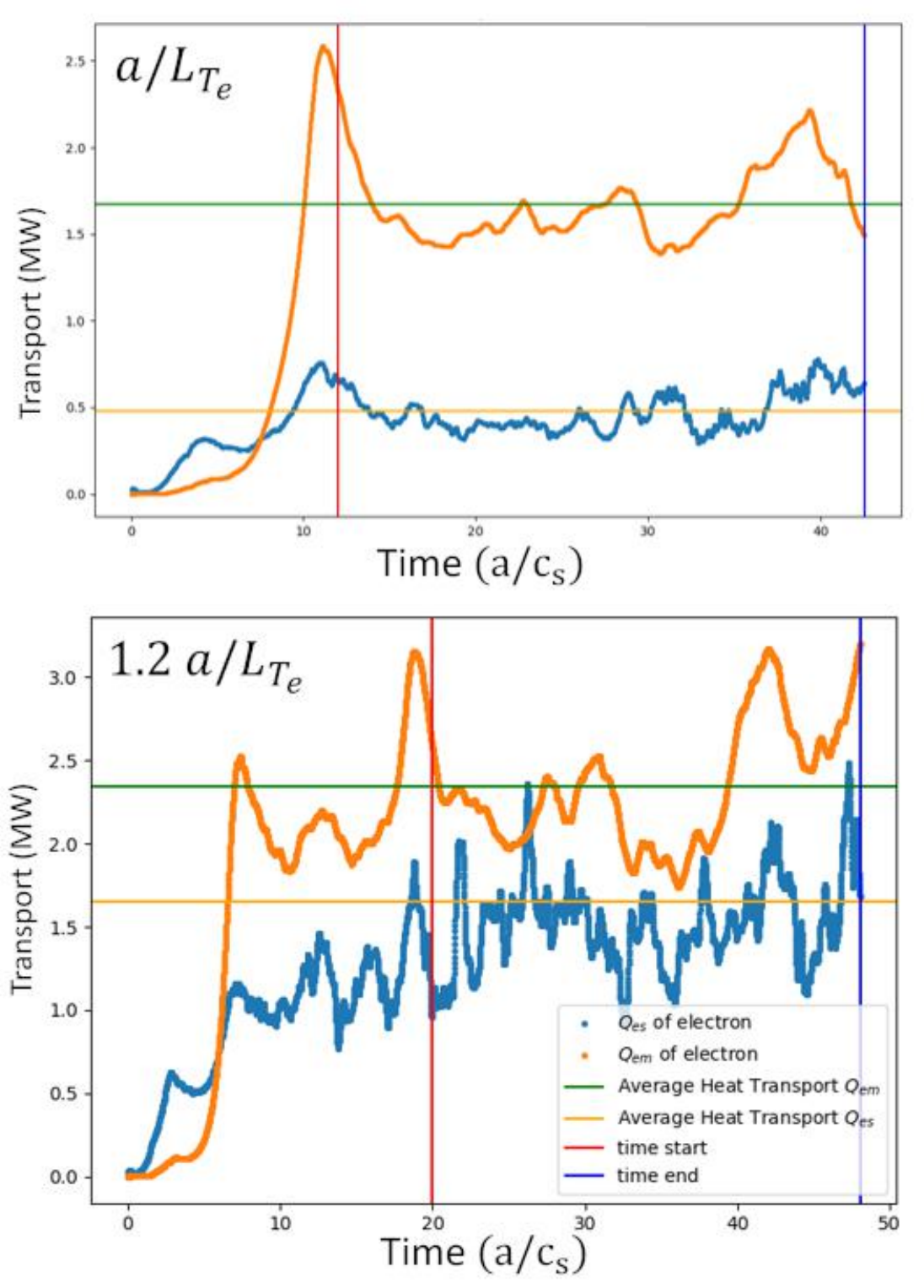}
        \centering
        \caption[font=5]{Local nonlinear simulations of MTM: the top plot is simulations using the nominal profile with saturated total electron heat transport of about 2.3MW; the bottom plot is simulations using the $120\%$ nominal $a/L_{T_e}$ with saturated total electron heat transport of about 3.99MW. The red vertical lines show the start of the saturated period, and the blue vertical lines mark the end of the saturated period. 
        The electromagnetic portions of the electron heat transports are represented by the orange lines, and their averages are represented by green horizontal lines. The electrostatic portions are represented by the blue lines, and their averages are represented by orange horizontal lines. }
        \label{fig:time_trace}
\end{figure}

The simulations have $k_y\rho_s$ ranging from 0 to 0.72. 
Simulations employ the three dynamic particle species (electron, deuterium, and carbon). They are based on the nominal equilibrium profile, and a profile with 120\% of the electron temperature gradient at the pedestal. The increase of the electron temperature gradient is intended to increase the transport level of MTM using profile variation within the experimental uncertainty, as MTM is driven by the electron temperature gradient.  

These simulations do not include background $E \times B$ shear since it stabilized the MTM turbulence when the shear is at significant levels.  We justify neglecting $E \times B$ shear by noting that there is a zero in the $E \times B$ shear rate in the middle of the pedestal at $\rho_{tor} \sim 0.983$, which is quite close to the flux surface at which we carry out our simulations $\rho_{tor} = 0.98$.  The $E\times B$ shear rate is highly uncertain as it is effectively the second derivative of impurity (carbon) pedestal profiles.  Based on the $E \times B$ shear profile, we expect anything between $\gamma_{ExB}(c_s/a) \sim 0-1$ is plausible within experimental uncertainties at $\rho_{tor}=0.98$.  Since the MTM is a rather localized fluctuation, it can comfortably live in this region of low $E \times B$ shear.  

Fig.~\ref{fig:time_trace} shows time traces of heat fluxes for two simulations: one with the nominal electron temperature gradient and another with a 20\% increase in an electron temperature gradient. The simulation with the nominal profile produces electromagnetic electron heat transport $Q_{EM} S=1.67\pm0.21MW$, electrostatic electron heat transport $Q_{ES} S=0.48\pm0.11MW$, in total produces $Q_{e} S=2.15\pm0.32MW$. Where $S$ is the area of the flux surface for the calculation of the transport based on the flux. The simulation with the  20\% increase in electron temperature gradient produces electromagnetic electron heat transport $Q_{EM} S=2.34\pm0.27MW$, electrostatic electron heat transport $Q_{ES} S=1.65\pm0.35MW$, in total produces $Q_{e} S=3.99\pm0.63MW$. 
As can be seen in Fig.~\ref{fig:time_trace}, the electromagnetic component of the transport saturates for a considerable length of time before the electrostatic component grows uncontrollably with the nonphysical transport level (just beyond the plot range). 

\subsubsection{Construction of synthetic diagnostics}

\begin{figure*}
        \includegraphics[width=0.9\textwidth]{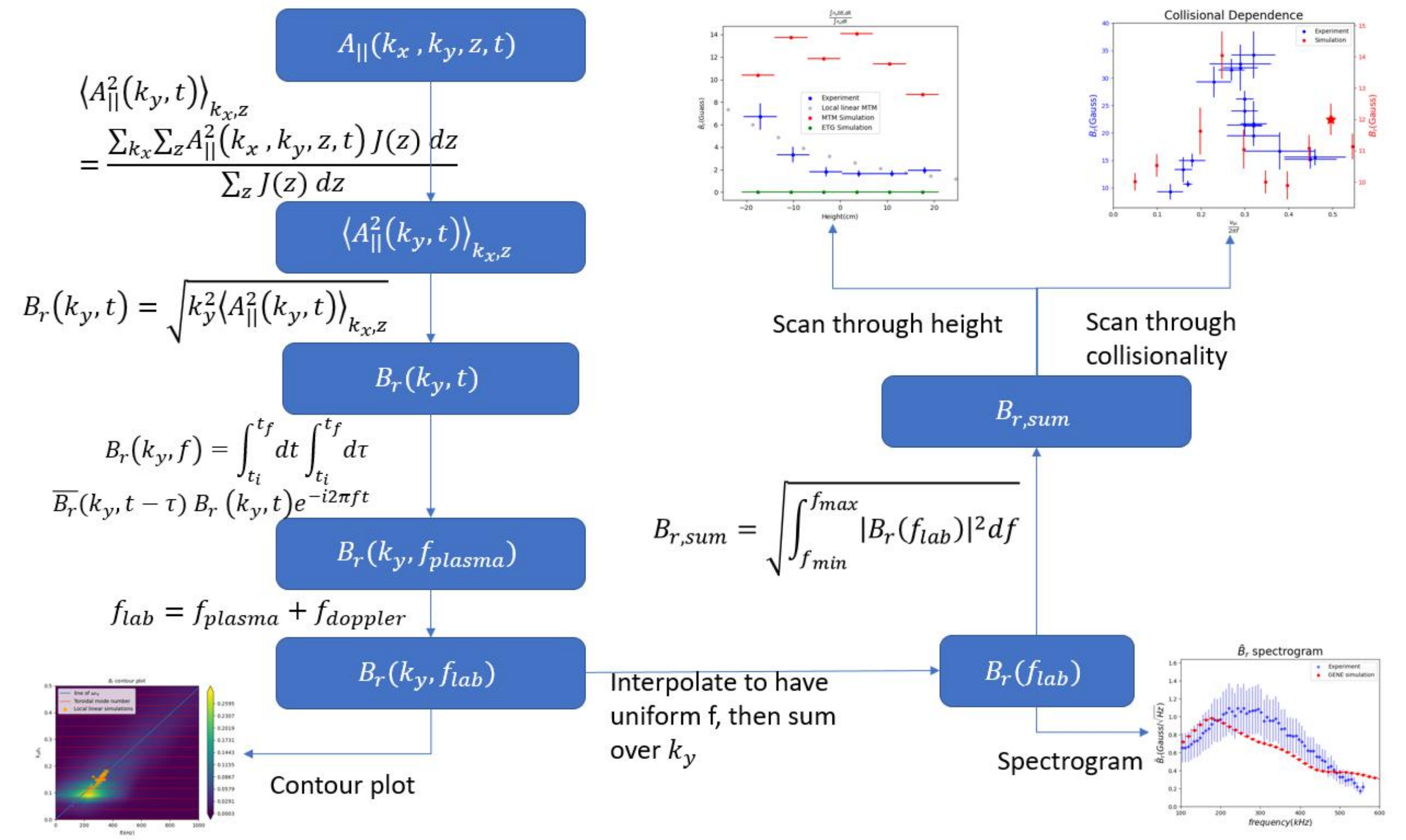}
        \centering
        \caption[font=5]{Processes of the synthetic diagnostic tools being constructed}
        \label{fig:syth}
\end{figure*}

Local nonlinear GENE simulations produce fluctuation data (electrostatic and magnetic vector potentials, density, temperature fluctuations, etc.) in a three-dimensional flux tube (a domain that follows the magnetic field lines, is extended in the parallel direction, and is limited in the perpendicular directions).  This data is processed in various ways for comparison with experimental fluctuation data, as illustrated in the flowchart in Fig.~\ref{fig:syth}.  



Fig.~\ref{fig:syth} shows the processes of the simulation diagnostic tools being constructed to directly compare with the experimental results. The diagnostic takes $A_{\|}(k_x,k_y,z,t)$ from GENE simulations, then converts to $B_r(k_x,k_y,z,t)$ in Gaussian units. 
The magnetic field is then summed over the $k_x$:  $B_r(k_y,z,t)=\sum_{k_x}B_r(k_x,k_y,z,t)$. To mimic the RIP diagnostic, the synthetic diagnostic tool does a Jacobian-weighted sum over a height of 7cm around the height of interest: $B_r(k_y,t)=\frac{\int B_r(k_y,z,t) J(z) dz}{\int J(z) dz}$, where $J(z)$ is the Jacobian that is a function of $z$. The coordinate of the simulation is non-Cartesian. In order to have a normalized sum over its Cartesian area, we employ the Jacobian-weight sum. The auto-correlated Fourier transform was performed to translate the information from time to frequency space:  $B_{r}\left(k_{y}, f_{plasma}\right)^2=\int_{t_i}^{t_f} d \tau \int_{t_i}^{t_f} d t \overline{B_{r}}\left(k_{y}, t-\tau\right) B_{r}\left(k_{y}, t\right) e^{-i 2 \pi f t}$. In this analysis, $t_i$ is the starting time, $t_f$ is the ending time,  $B_{r}\left(k_{y}, f_{plasma}\right)$ is in units of $Gauss/\sqrt{kHz}$ as calculated spectral density. The Doppler shift is also added to boost the frequency from the plasma frame to the lab frame: $f_{lab}=f_{plasma}+ f_{Doppler}$, where the $f_{Doppler}=1.5 kHz$ for toroidal mode number n=1 (for more details about the Doppler shift, refer to appendix section). The contour plot in terms of $B_r$ across $k_y\rho_s$ and $f_{lab}$ then can be plotted from $B_{r}(k_y,f_{lab})$. Interpolation will be conducted to unify the frequency space in preparation for the summation over $k_y$ to arrive at the final spectrogram $B_{r}(f_{lab})$. To sum over a bandwidth of frequency the formula is the following: $B_{r,sum}=\sqrt{\int_{f_{min}}^{f_{max}}|B_r(f_{lab})|^2df}$ The following calculation of $B_r$ as a function of height will be summed around the mid-plane, and integrated over the frequency of 150kHz to 500kHz to match the integration range from the experiment\cite{RIP_Chen_POP_2021}
The collisional dependence calculation is integrated over the full frequency range at the mid-plane to match with the method of data processing in the experiment\cite{RIP_Chen_POP_2020,RIP_Chen_POP_2021}.  All analysis is performed during the intermediate time period when the transport is in a saturated state (the starting time and ending time are marked as red and blue vertical lines respectively in Fig.~\ref{fig:time_trace}).

\subsubsection{Frequency Spectra and Fluctuation Amplitudes}\label{sec:NL_freq}

\begin{figure}[ht]
        \includegraphics[width=0.45\textwidth]{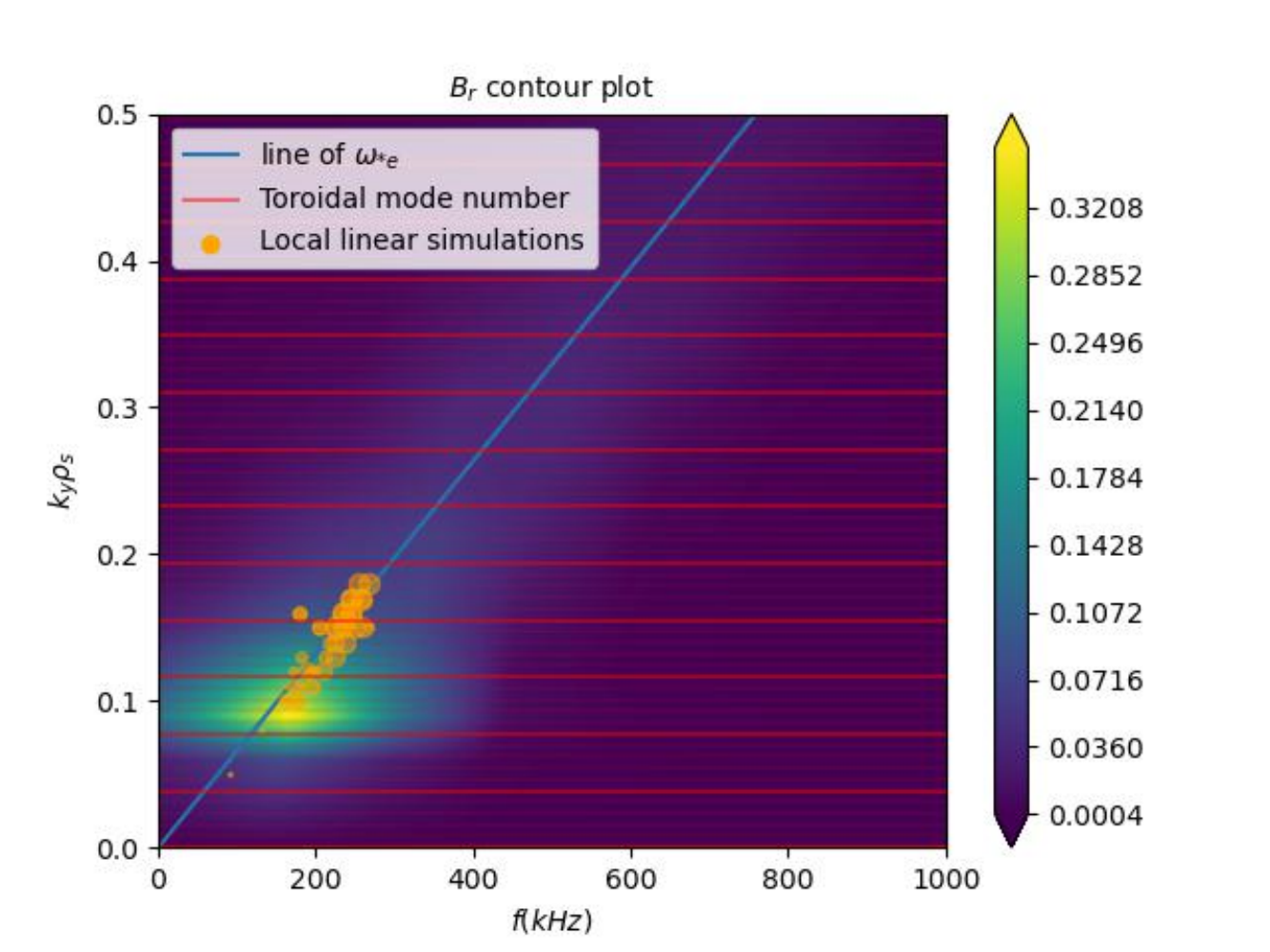}
        \centering
        \caption[font=5]{This plot shows the comparison between local nonlinear simulations (contour plot), local linear simulations (orange dots), and theory (blue line). The contour plot shows $B_r$ as a function of frequency and $k_y \rho_s$ calculated from local nonlinear simulations, where $B_r$ has the unit of $Gauss/\sqrt{Hz}$. The orange dots are the results of the local linear simulations with different $k_x\rho_s$ and $k_y\rho_s$. The size of the orange dots represents the growth rate of each mode. The blue line shows the MTM frequency expected from theory\cite{Kotschenreuther_2019}: electron diamagnetic frequency in the lab frame. 
        The red horizontal lines correspond to the toroidal mode numbers, starting from n=0, the darker solid red lines have an increment of 5 for toroidal mode number n. }
        \label{fig:NL_heat}
\end{figure}

Building upon the agreement between linear growth rates/frequencies and the RIP frequency spectrum in Fig.~\ref{fig:LL_exp}, here we analyze frequency spectra from the nonlinear simulations.  The contour plot of $B_r$ as a function of mode numbers and frequency is shown in Fig.~\ref{fig:NL_heat}. The fluctuations are maximum at n=12, in good agreement with the experimental expectation of toroidal mode numbers in the range ($10-20$) as noted in Ref.~\cite{RIP_Chen_POP_2021}. For reference, a line representing $\omega_{*e} = k_y \rho_s c_s (1/L_{T_e} + 1/L_{n_e})$ is also shown in the figure along with the frequencies of the linear eigenmodes.  All quantities---from basic theory, linear simulation, and nonlinear simulation---align nicely.  

\begin{figure}[ht]
        \includegraphics[width=0.45\textwidth]{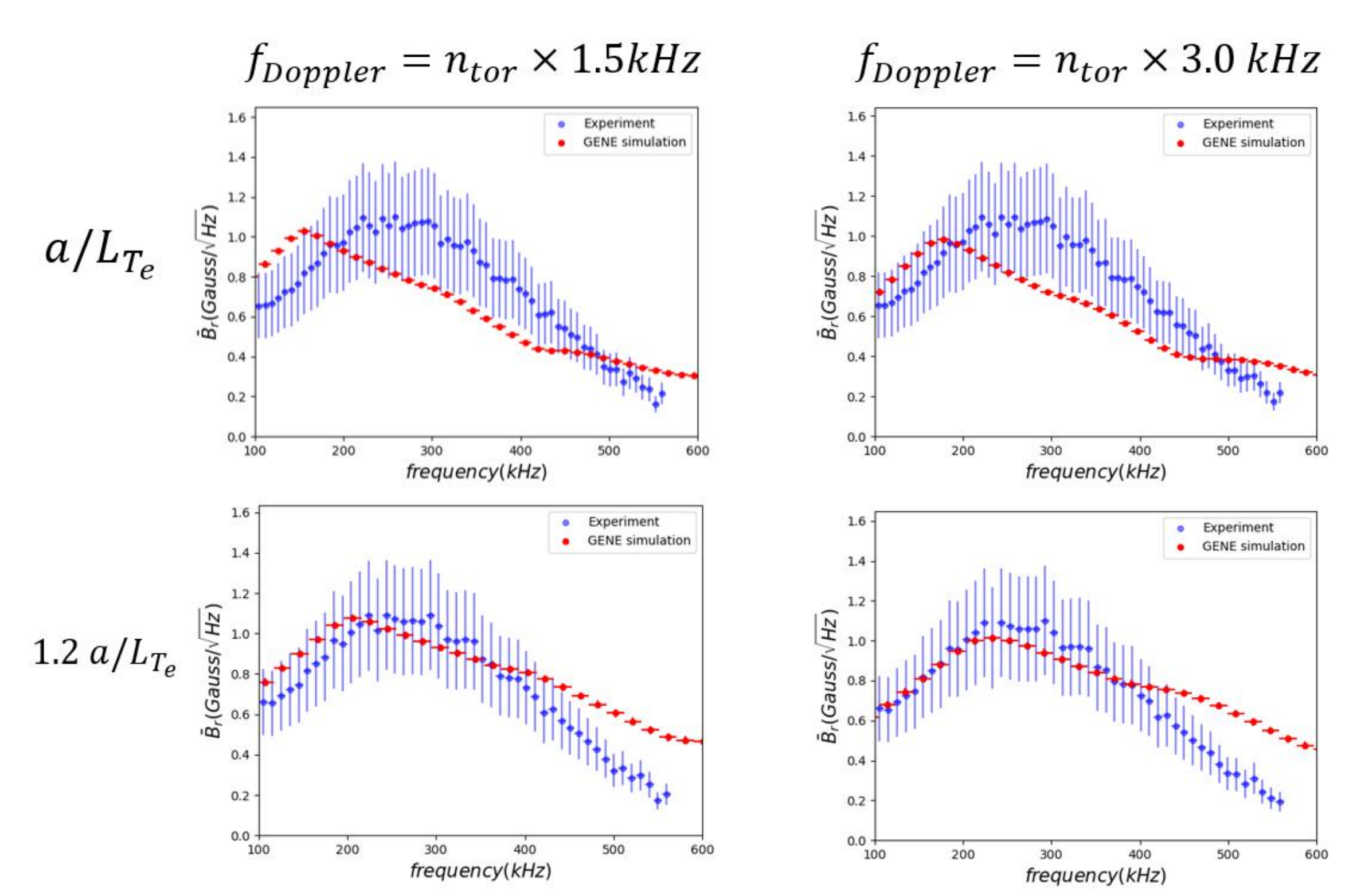}
        \centering
        \caption[font=5]{The comparison of frequency of the experiment 175823 spectrogram (blue) and the local nonlinear simulations (red) where the top two plots show the simulation with nominal profiles and the bottom two plots show the simulation with $120\%$ of $a/L_{T_e}$. The left two plots are using $1.5kHz$ of Doppler shift for toroidal mode number $n_{tor}=1$, while the right two plots are using $3.0kHz$}
        \label{fig:NL_f}
\end{figure}

 Fig.~\ref{fig:NL_f} shows the spectral density of magnetic fluctuations as diagnosed by both RIP measurement (blue) and the GENE local nonlinear simulations (red).  The experiment has $25\%$ of the uncertainty of the magnetic fluctuation measurement which is represented as the vertical bar. 
In this figure, we show the frequency spectra for different simulation scenarios representing various uncertainties in the electron temperature gradient and the Doppler shift: nominal gradients and Doppler shift (top left); nominal gradients with the increased Doppler shift (top right---recall appendix section for a discussion of the Doppler shift);  electron temperature gradient increased by 20\% with nominal Doppler shift (bottom left); and the increased temperature gradient and increased Doppler shift (bottom right).  The simulation with the increased temperature gradient agrees very well with the peak and the general shape of the frequency spectrum.  This substantial shift to higher frequencies for the increased gradient case is attributable to two factors: (1) a slight increase in $\omega_{*e}$, and (2) a shift
of the nonlinear fluctuation spectrum to higher $k_y$ (and higher frequency). In order to be consistent with the experimental calculation for the spectrum, both simulations and experiments use Welch's method\cite{Welch} with 'hann' window. 


The apparent good agreement between simulation and experiment in Fig.~\ref{fig:NL_f} is, unfortunately, a coincidence, since the experimental data is averaged over the entire radius of the device (including regions in the core with, presumably, much lower magnetic fluctuation levels).  In contrast, the GENE data is averaged over only the small flux-tube radial domain localized where the MTM fluctuations are most potent.

We may speculate on various reasons for the discrepancy in fluctuation amplitude.  First, the recent RIP diagnostic upgrade  \cite{doi:10.1063/5.0040306} provides lower noise floor measurement. Interestingly, the measured broadband magnetic fluctuation amplitude in discharges similar to 175823 has been reduced since then. An example can be seen in Chen (2022) \cite{Chen_RIP_U_2022}.   In shot 183225, a shot not identical but similar to 175823 (shape, q95, density, etc.), measured line-averaged broadband magnetic fluctuation amplitude (150-500kHz) is only 3-4 Gauss, compared to 15 Gauss measured in 175823. It appears that a higher noise floor in FY18 might have introduced an uncertainty leading to an overestimate of absolute magnetic fluctuation amplitude. This potentially could explain part of the discrepancy. 
Second, as noted above, these local nonlinear simulations have limitations that may produce discrepancies with the experimental observations.  
Third, it may point to the need to construct more sophisticated synthetic diagnostics.
Another possible source of the discrepancy can be identified in the poloidal dependence of the fluctuations.  This can be seen in Fig.~\ref{fig:MTM_Apar_mode}, which shows $A_{||}$ as a function of the parallel direction (parameterized by the poloidal angle, $\theta$).  Note that the amplitude peaks strongly away from the outboard mid-plane where the diagnostics are centered.  In fact, the flux and fluctuation amplitudes are very low at this location.  Consequently, the simulated amplitudes at this location are perhaps a less-robust reflection of the quality of the simulation in comparison with the experiment.  Stated differently, the peak amplitudes are the clearest reflection of the thermodynamic drives of the system and would thus be expected to be the most reliable signals of the simulation.  This is reinforced by the fact that the simulations produce realistic transport levels---2.1MW to 4.2MW of electromagnetic heat flux, which is closely connected to magnetic fluctuation levels~\cite{Pueschel_2008} (the best estimates of the experimental electron heat losses range from $2.3-4.3$ MW as discussed below).  Consequently, we view this hypothesis to be very plausible.

\begin{figure}[ht]
        \includegraphics[width=0.45\textwidth]{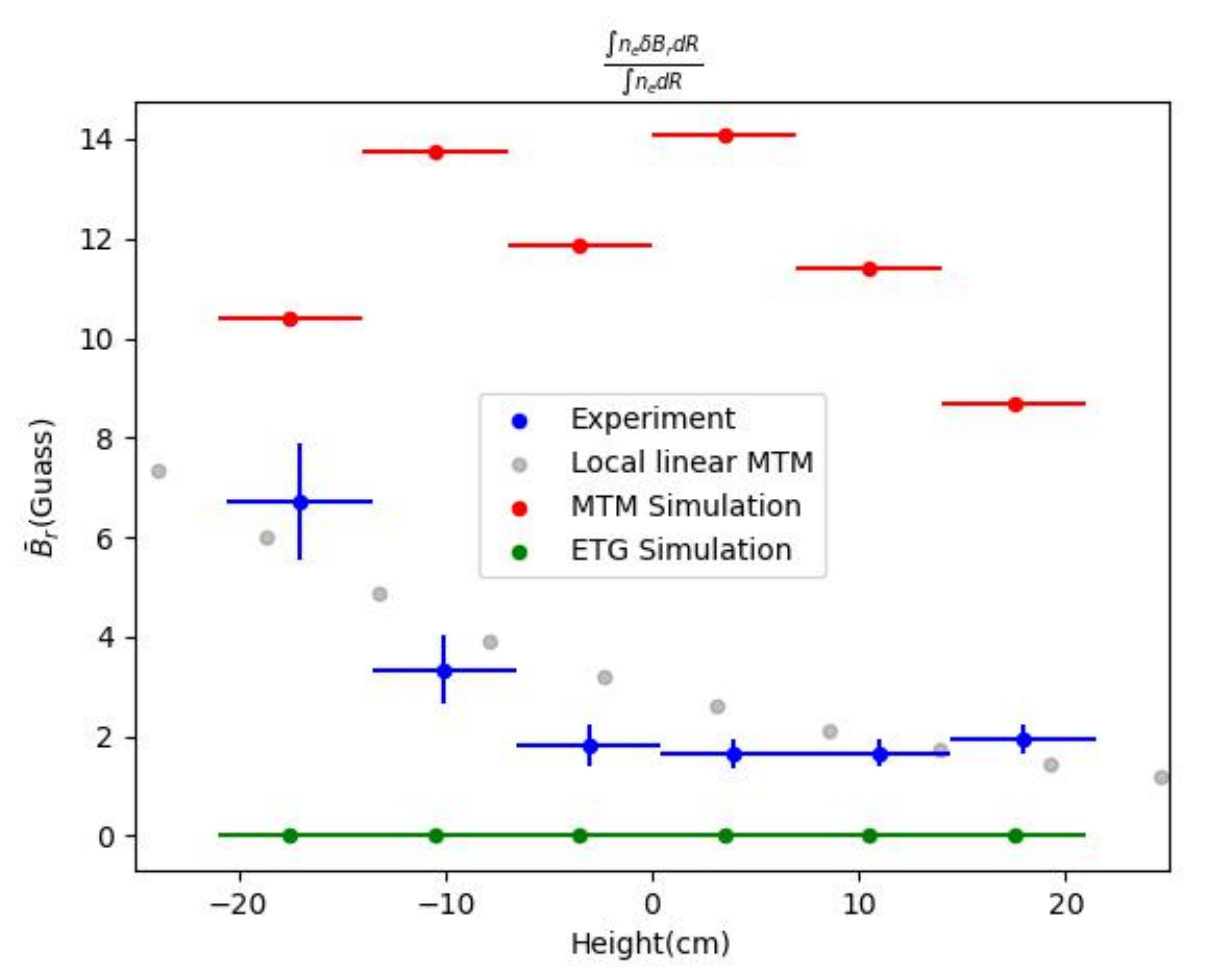}
        \centering
        \caption[font=5]{Comparison between the experiment discharge 179451 and simulations of $B_r$ over height around the mid-plane. The blue dots are experiment observation, the red dots are the local nonlinear simulation for MTM, the green dots are the local nonlinear simulation for ETG, the grey dots are the local linear simulation for MTM. The local linear $B_r$ has been normalized in order to be compared with other quantities. }
        \label{fig:Br_Z}
\end{figure}

Similar considerations should be kept in mind when comparing other features of the simulations.  The RIP diagnostic was used to measure magnetic fluctuations at multiple vertical locations (recall Fig.~\ref{fig:RIP}).  The magnetic fluctuation amplitude strongly decreases in the vertical direction with no apparent symmetry around the outboard mid-plane as shown in Fig.~\ref{fig:Br_Z} for discharge 179451.  In order to compare the spatial dependence between experiments and simulations, we integrate over the peak frequencies (150kHz to 500kHz)  and extract the fluctuation amplitude ($\delta B_r=\sqrt{\int^{500kHz}_{150kHz} B_r(f)^2 df}$) at various vertical locations.  The simulated fluctuation amplitudes are also shown in Fig.~\ref{fig:Br_Z}.  Due to the fact that the simulation is based on inputs from a different discharge (175823), only qualitative comparisons should be made.  The nonlinear MTM simulation (red) exhibits a relatively constant amplitude in the vertical direction.  Interestingly, the linear eigenmode (shown with grey symbols) exhibits a vertical dependence closely matching that of the experimental observation.     

For reference, we also plot the magnetic fluctuation amplitude extracted from a nonlinear ETG simulation (green) performed at the same radial location.  This fluctuation amplitude is negligible: $\delta B_r =10^{-3} Gauss$ despite the fact that the ETG fluctuations do peak at the outboard midplane. This is a clear observation distinguishing MTM and ETG fluctuations, which otherwise have many `fingerprints' in common (i.e., they are driven by electron temperature gradients, have negative frequencies, and produce almost exclusively electron heat transport).


\subsubsection{Collisionality Dependence}

\begin{figure}[ht]
        \includegraphics[width=0.45\textwidth]{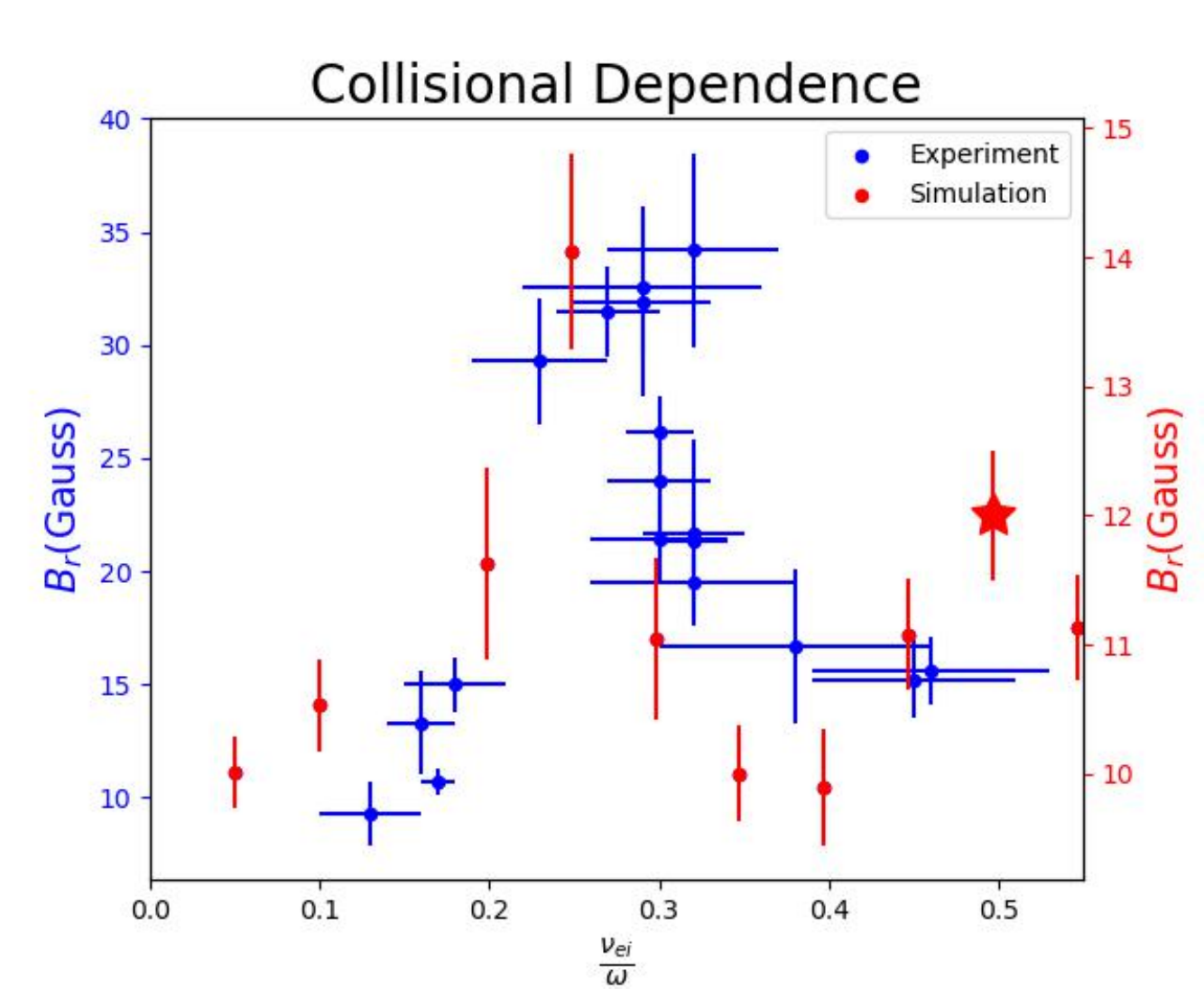}
        \centering
        \caption[font=5]{Collision dependent comparison between experiments (blue dots with blue axis) and local nonlinear simulations (red dots with red y axis), the star is denoted as the simulation with the nominal value taken from discharge 175823. $\nu_{ei}$ is the electron-ion collision frequency, $\omega$ is the mode frequency. It is important to note that the experiment and simulations $B_r$ is plotted on a different scale.  This plot is intended to show the qualitative comparison between experiments and simulations. }
        \label{fig:nu_ei}
\end{figure}

Non-monotonic collisionality dependence is a hallmark of microtearing modes. The collision frequency at which the MTM peaks is sensitively dependent on a multitude of variables---$\eta\equiv L_n/L_T$, magnetic shear, $k_y$~\cite{Joel_prl,gladd}. 
In Ref.~\cite{RIP_Chen_POP_2021}, a set of discharges (175816, 175821, 175823, 176303, 176462, 176463, 176464) was analyzed in order to survey the collisionality dependence of pedestal magnetic fluctuations.
This set of discharges varies collisionality while minimizing variation in other relevant parameters. The fluctuation amplitudes from the RIP diagnostic and local nonlinear simulations with different collisionality are shown in Fig.~\ref{fig:nu_ei}, qualitatively reproducing the non-monotonic behavior expected from MTMs. The amplitudes are normalized to their respective peak values. 

Notably, the simulations and experiment find good agreement in the value of collisionality at which the fluctuation amplitudes peak.  They also exhibit reasonable agreement in the breadth of the distribution, although the experimental results are somewhat narrower in collision frequency.  A similar (preliminary) comparison---between the experimental fluctuation amplitudes and simulated linear growth rates---was shown in Ref.~\cite{RIP_Chen_POP_2020}.  The comparison with nonlinear simulations shown here is much more compelling---both the peak collisionality and the breadth of the distribution are in much better agreement, pointing to key nonlinear modifications to the linear picture. The sharper falloff of the nonlinear simulations than linear as collisionality increases may be attributed to the ESDW becoming increasingly dominant, consistent with the dependencies shown in Fig.~\ref{fig:EV_nu}. 

\subsubsection{Fluctuation Ratio}

In addition to the fluctuation amplitudes, fluctuation ratios are highly relevant since they can (1) be compared to linear eigenmodes, and (2) often clearly distinguish between different classes of modes.

BES \cite{bes} can measure $\delta n_e/n_e$ at the outboard mid-plane, radially localized in the pedestal. The calculation of $\delta n_e$ from local nonlinear simulations is done in the same way as $\delta B_r$ calculations except the quantities being used -- one uses $\delta n_e$ the other used $\delta B_r$. With this, one can calculate the fluctuation ratio $\frac{\delta B_r/B_0}{\delta n_e/n_e}$. For the discharge of interest (175823) at the mid-plane, this is measured to be $\frac{\delta B/B_0}{\delta n_e/n_e}= 0.08\pm 0.03$ (150kHz $\sim$ 500kHz)~\cite{RIP_Chen_POP_2020}. 

This can be directly compared with simulation results.  For MTM, $\delta B_r=12Gauss$, $\delta n_e=6.16\times 10^{16}/m^3$; for ETG $\delta B_r=0.067Gauss$, $\delta n_e=3.6\times 10^{16}/m^3$. Recall from the Table \ref{ch:sim_para}, $B_0=19773Gauss$, $n_e=3.57\times 10^{19}/m^3$.  Combining these calculations, we arrive at estimates from the local nonlinear MTM simulations of $\frac{\delta B/B_0}{\delta n_e/n_e}= 0.36$. 
On the other hand, ETG has the fluctuation ratio of $\frac{\delta B/B_0}{\delta n_e/n_e}= 0.0038 $. Therefore, the ratios calculated from MTM, ETG, experiment have the following relation.  

\begin{equation}
   \text{ratio}_{ETG}\ll \text{ratio}_{experiment} \ll \text{ratio}_{MTM} 
\end{equation}

The ratio from the MTM simulation is over a factor of 4 greater than the experiment. There are a few possibilities for such a discrepancy. It is possible that other electrostatic instabilities are also present but not captured in the scale ranges of our simulations.  Another possibility is that KBM fluctuations are present in the discharge but not captured in the simulations (Ref.~\cite{Halfmoon_MTM} investigates such possibilities for another DIII-D discharge).  Such fluctuations would produce relatively stronger density fluctuations while the MTM is the primary contributor to the magnetic fluctuations. We also note once again the fact that the outboard mid-plane is a region of low simulated fluctuation levels, and so maybe a challenging location for comparisons.  

One clear conclusion, however, is that ETG fluctuations are very unlikely to be responsible for the observed magnetic fluctuations, since $\frac{\text{ratio}_{ETG}}{\text{ratio}_{experiment}} =0.0475 $. Such ratios can be used broadly for experiments and simulations to determine the types of instabilities\cite{Kotschenreuther_2019}. 

\subsubsection{Transport}

\begin{figure}[ht]
        \includegraphics[width=0.45\textwidth]{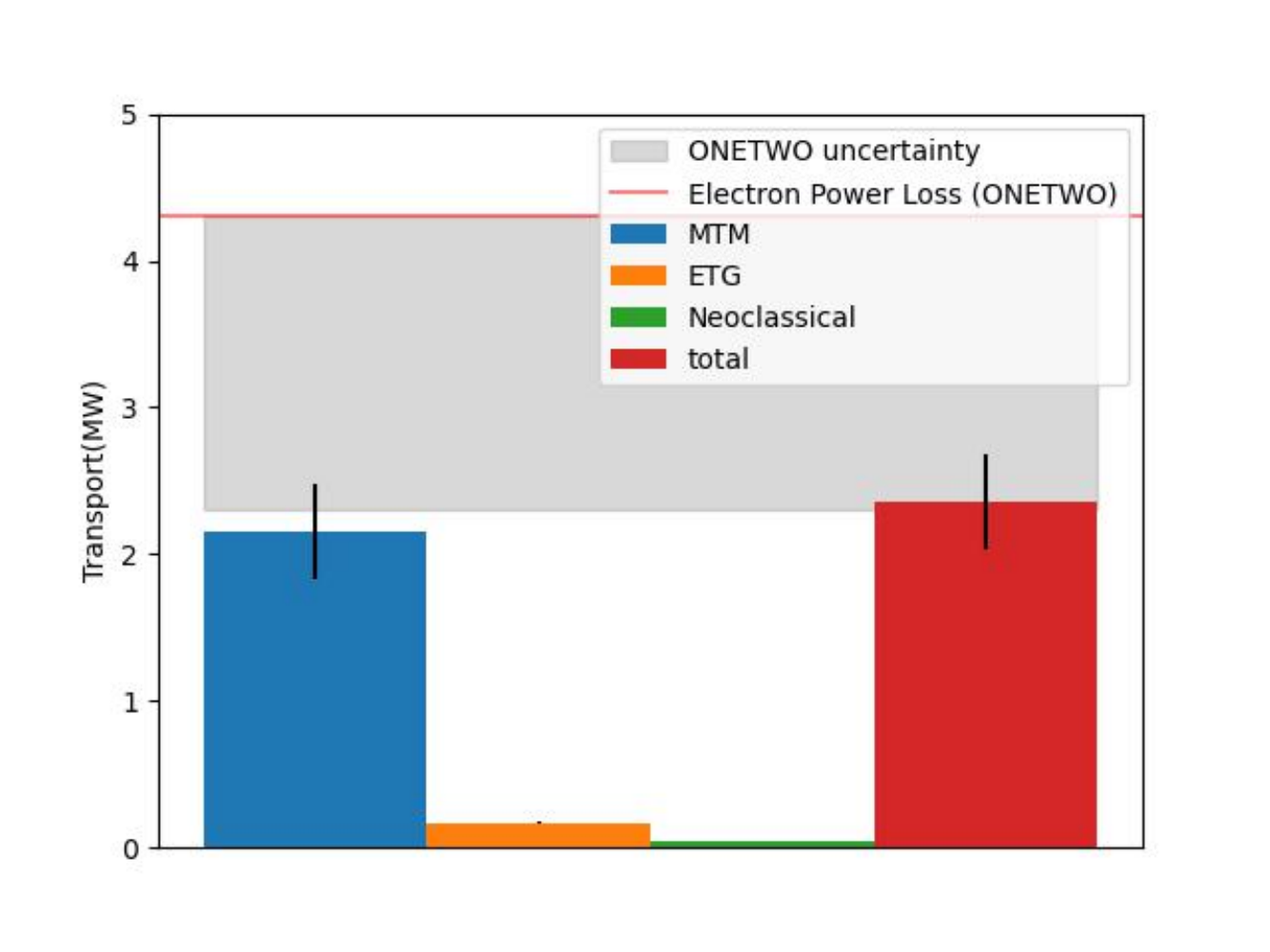}
        \centering
        \caption[font=5]{Power balance: the electron power loss calculated from the ONETWO is $4.3MW$ (red horizontal line). The  grey transparent rectangle represents the uncertainty of ONETWO estimation
        . The simulation calculated transport from MTM (blue bar) is $2.15\pm0.32MW$, ETG (orange bar) is $0.17\pm0.01MW$. neoclassical (green bar) is $0.037MW$. The total simulated electron heat transport added up to $2.36\pm0.32MW$ which is marked as the red bar}
        \label{fig:power}
\end{figure}

\begin{figure}[ht]
        \includegraphics[width=0.45\textwidth]{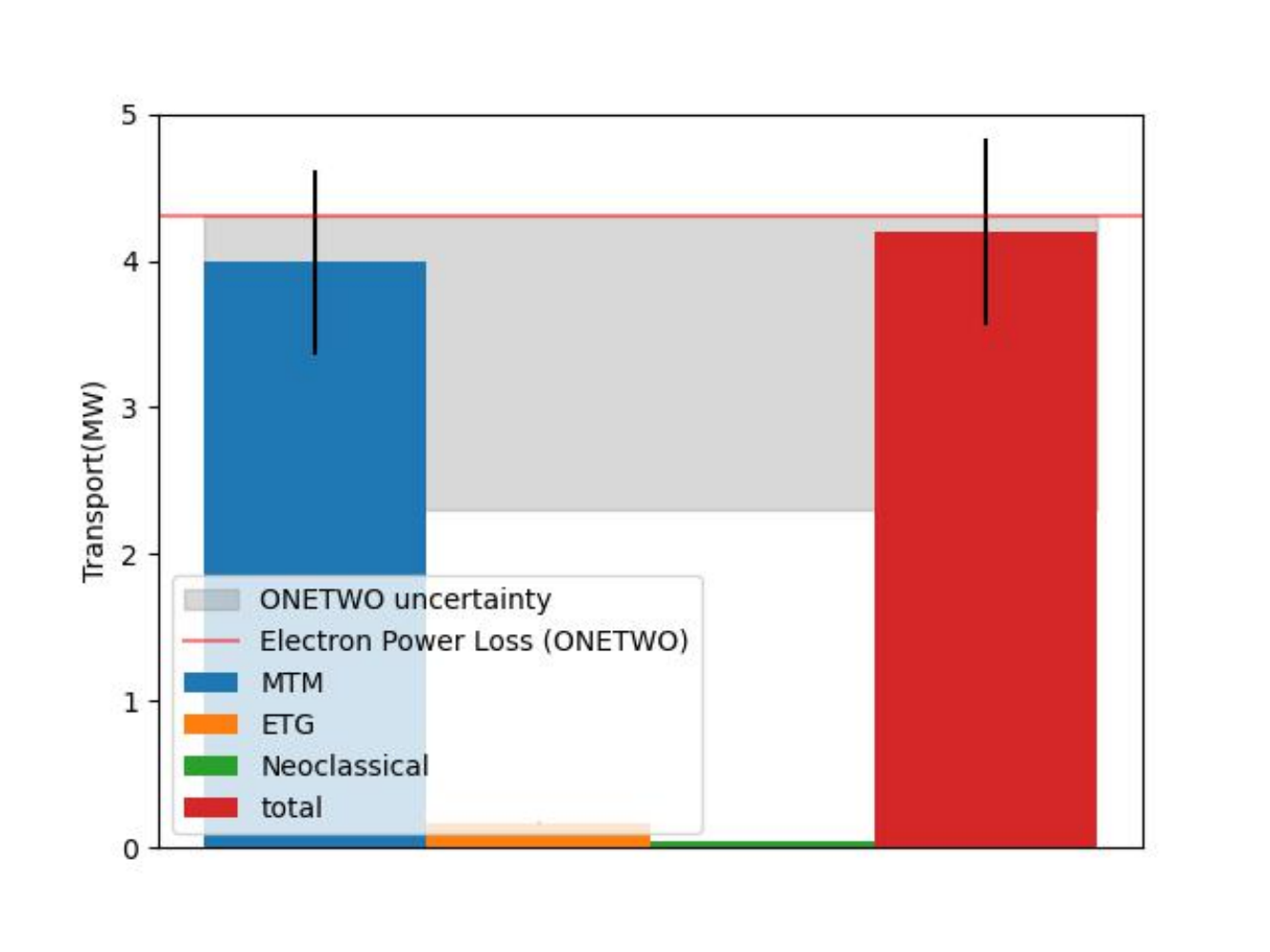}
        \centering
        \caption[font=5]{Power balance: the electron power loss calculated from the ONETWO is $4.3MW$ (red horizontal line). The  grey transparent rectangle represents the uncertainty of ONETWO estimation
        . The simulation (with $120\%$ of the nominal electron temperature gradient) calculated transport from MTM (blue bar) is $3.99\pm0.63MW$, ETG (orange bar) is $0.17\pm0.01MW$,  neoclassical (green bar) is $0.037MW$. The total simulated electron heat transport added up to $4.20\pm0.63MW$ which is marked as the red bar}
        \label{fig:power_120}
\end{figure}

The experimental estimate of the total electron heat transport for discharge 175823 is $\sim 2.3-4.3MW$. The ONETWO analysis uses carbon ion temperature. However, the recent study 
\cite{Haskey_2022} shows that using carbon ion temperature may underestimate the power loss in the ion channel and therefore overestimate the electron power loss. The real power loss due to electron could be roughly 2MW lower which is a grey box marked for ONETWO uncertainty in Fig.~\ref{fig:power} and Fig.~\ref{fig:power_120}.

Fig.~\ref{fig:power} shows the contributions from the nonlinear ETG and MTM simulations.  For the nominal scenario, these two components of the transport reach the lower band of the ONETWO uncertainty estimate.  If we consider also the MTM simulation with a higher temperature gradient, the simulations lie more clearly in the expected range. Fig.~\ref{fig:power_120} shows the transport with MTM simulations with $120\%$ of the electron temperature gradient.  In light of the fact that magnetic fluctuation levels are closely connected to transport levels, this may suggest that the peak magnetic fluctuation amplitudes accurately reflect those in the experiment while the simulated levels at the outboard midplane are not as accurate.   

Despite the aforementioned long-time runaway of the simulations, the fact that the intermediate saturation period produces reasonable transport levels adds some credence to the simulations.


It is also worth mentioning that local nonlinear simulation yield total ion transport of $Q_i S = 0.15 \pm 0.13MW$ with electrostatic ion heat flux is $Q_{es,i} S = 0.15 \pm 0.13M W$ , and electromagnetic ion heat flux $Q_{es,i} S = 0.005 \pm 0.003MW$. The transport coefficients ratio can be used as a fingerprint \cite{Kotschenreuther_2019} to identify the modes of instabilities. Table \ref{ch:ion_NL_coeff} shows such ratios for the ion scale local nonlinear simulations. And they point toward MTM. 

\begin{table}[h]
\centering
\begin{tabular}{|l|l|l|}
\hline
$Q_{em,e}/Q_{es,e}$ & $Q_i/Q_e$ & $D_e/\chi_e$ \\ \hline
2.56                & 0.13      & 0.043       \\ \hline
\end{tabular}
\caption[font=5]{Transport coefficients ratios for the ion scale local nonlinear simulations, where $Q_{em,e}$ is electron electromagnetic heat flux, $Q_{em,e}$ is electron electrostatic heat flux, $Q_{e}$ is electron heat flux, $Q_{i}$ is ion heat flux, $D_e$ is electron particle transport coefficient, $D_e$ is electron heat transport coefficient. }
\label{ch:ion_NL_coeff}
\end{table}

A neoclassical transport calculation using NEO~\cite{Belli_2008,Belli_2009,Belli_2011} was conducted at $\rho_{tor}=0.98$ with electron heat transport of $Q_e S=0.037MW$ and ion heat transport of $Q_i S=1.33MW$.  

In short, the combination of MTM and neoclassical transport appears to account for most of the transport in the electron and ion heat channels, respectively, for this discharge.  Clearly other transport mechanisms are active in other scenarios and discharges (and possibly in this discharge as well).  For example, ETG transport has been shown to be significant in the pedestal in several studies~\cite{Hatch_2015,Hatch_2017,MTM_power,osti_1776857,chapman2022}.  Interestingly, Ref.~\cite{osti_1776857} finds substantial ETG transport, but at levels insufficient to account for power balance (Ref.~\cite{Ehab_MTM} showed that MTM likely accounts for the remaining transport).  Moreover, ion scale electrostatic transport has been shown to be significant in certain scenarios as well~\cite{Hatch_2017,MTM_power,Haskey_2022}.  Ref.~\cite{Haskey_2022} finds substantial ion scale turbulent transport in the low collisionality regime where neoclassical is insufficient to account for power balance.  The analysis was aided by innovative measurements of the main ion temperature profile, which exhibited substantially steeper gradients than those inferred from the C profile.  Unfortunately, such measurements were not available for our study.  Consequently, the possibility of ITG transport remains and open question for this discharge.



\section{Conclusions and Discussion}

Local linear and nonlinear gyrokinetic simulations of DIII-D discharge 175823 identify MTMs as the most likely source of the magnetic fluctuations measured by the RIP diagnostic.  Linear MTMs have growth rates that peak at frequencies closely matching the peak of the experimentally-diagnosed frequency spectrum.  Despite some numerical issues with nonlinear simulations, there exists an intermediate time range in which the simulations are well-behaved and produce turbulence that finds clear connections with the experimental observations.  Notably, nonlinear simulations produce frequency spectra of $\delta B_r$ that are in good agreement with the peak and shape of the diagnosed spectrum of the magnetic fluctuations.  Nonlinear simulations also reproduce the variation of magnetic fluctuation amplitudes observed in an experimental scan of collision frequency---both experiment and simulations agree on the value of collisionality at which the fluctuations peak.  The simulations also produce realistic levels of electron heat transport (2-4 MW within reasonable uncertainty bounds, which is precisely in the range of experimental expectation).  

The simulations do not find close agreement with the absolute magnetic fluctuation levels.  This may be attributable to deficiencies in the simulations (e.g., the local approximation produces artificial saturation levels), or the need for more sophisticated synthetic diagnostics.  However, we posit that the main cause of the discrepancy is the fact that the RIP diagnostic targets the midplane, where simulation identifies the fluctuation levels to be at a minimum.  We expect the peak fluctuation amplitudes further away from the midplane to be a much more robust quantity for comparison.  This hypothesis is reinforced by the observation that the simulations produce electromagnetic transport levels (which are closely connected to magnetic fluctuation levels) in good agreement with experimentally-expected electron heat losses.  

Comparisons between the ratio of magnetic to density fluctuations also do not find close agreement.  This may be due to fluctuations from other instabilities (KBM or some electrostatic mode) that are not captured in the scale ranges or parameter points we simulate.  However, the experimentally diagnosed fluctuation ratios are in such a range that typical electrostatic fluctuations can be eliminated as candidates.  

These results strongly support the conclusion that MTMs are very likely to be responsible for the observed magnetic fluctuations.  Figs.~\ref{fig:NL_heat} and ~\ref{fig:NL_f} are  especially relevant. Fig.~\ref{fig:NL_heat} demonstrates close agreement between basic theory, linear simulation results, and nonlinear simulation results, and Fig.~\ref{fig:NL_f} extends this agreement to the experimentally measured frequency spectrum for the magnetic fluctuations.   




\begin{table}[]
\begin{tabular}{|l|l|l|l|l|}
\hline
Quantity &Experiment & MTM  & ETG  \\ \hline
Frequency& 250kHz&250kHz& $>$500kHz \\ \hline
Direction  & Electron & Electron& Electron     \\ \hline
$\frac{\delta B_r/ B_0}{ \delta n_e/ n_e}$  &  $0.08\pm 0.03$ & 0.36 & 0.0038     \\ \hline
\begin{tabular}[c]{@{}l@{}}Non-monotonic\\ collisional \\ dependence\end{tabular} & Yes& Yes & No    \\ \hline
\begin{tabular}[c]{@{}l@{}}Electron Heat\\
Transport\end{tabular} & $2.3\sim4.3$MW  & $3.99$MW  & $0.17$MW  \\ \hline
\end{tabular}
\caption[font=5]{Comparisons of MTM, ETG simulations, and experiment observations}\label{ch:sum}
\end{table} 

As shown in Table \ref{ch:sum}, the results show a high level of agreement between GENE simulations for MTM and RIP observations. All the evidence gathered supports the idea that MTM is an important electron heat transport mechanism.

\begin{figure}[ht]
        \includegraphics[width=0.45\textwidth]{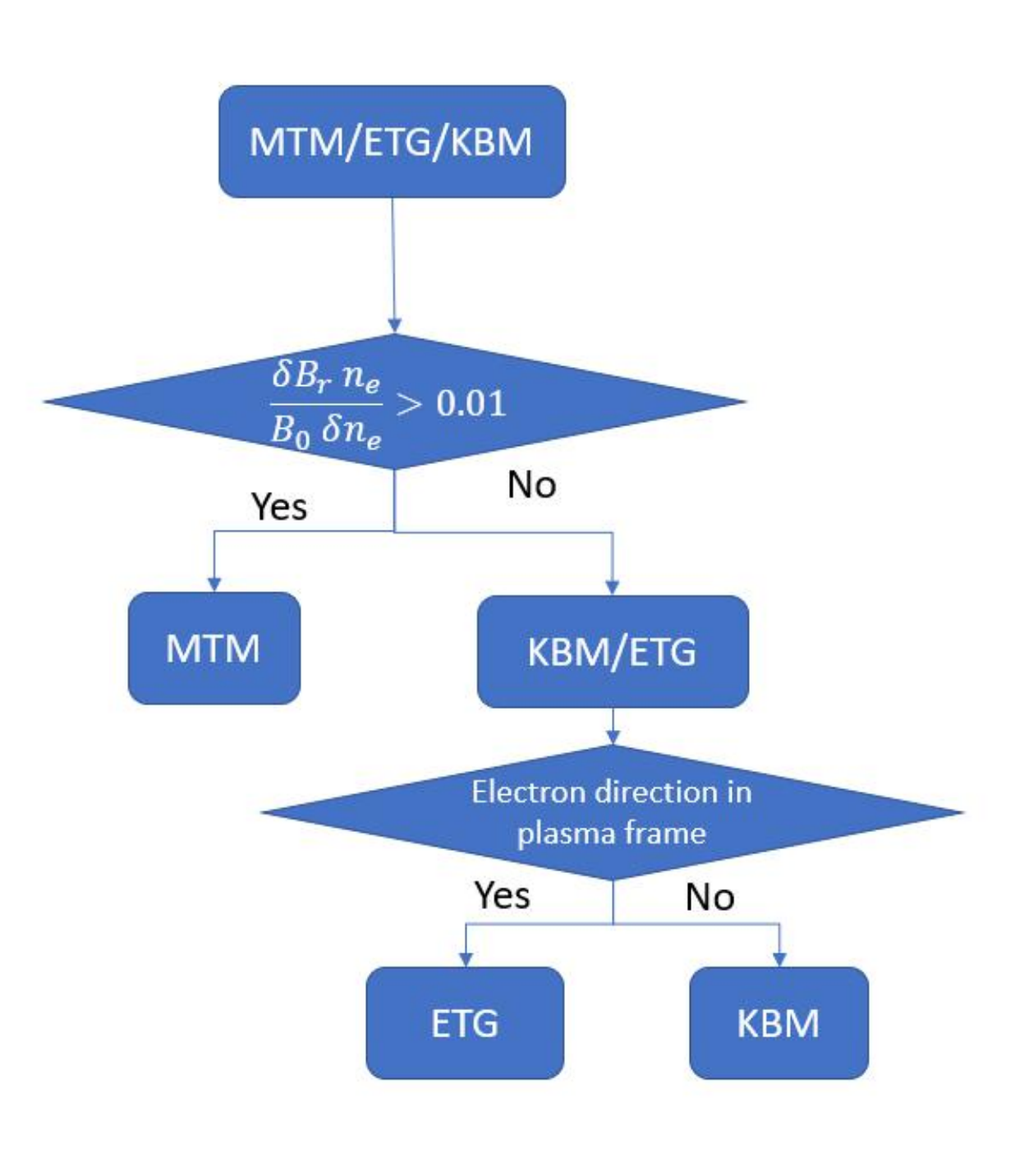}
        \centering
        \caption[font=5]{Flowchart for distinguishing the MTM, ETG and KBM.The top diamond is to judge if the instability is largely magnetic, which distinguish MTM from KBM and ETG. The second diamond is to judge if the frequency is in the electron direction which can be used to differ KBM and ETG. }
        \label{fig:flow}
\end{figure}

Inspired from the fingerprint method in Kotschenreuther (2019)\cite{Kotschenreuther_2019}, Fig.~\ref{fig:flow} shows, experimentally, one can exercise the following flowchart to separate MTM, ETG, and KBM using observation of frequency in plasma frame, magnetic fluctuations, electron density fluctuations: If the $\frac{\delta B_r n_e}{B_0 \delta n_e} >0.01$, then the instabilities is likely to be MTM \cite{Hillesheim_2015}. Otherwise, it could be KBM or ETG. If the frequency of the instability is in the electron direction in the plasma frame then the instability is likely to be ETG; otherwise, it will be KBM. If one applies such a flowchart to the results listed in the Table \ref{ch:sum}, it is rather easy to find that what the experiment observed is MTM.

\section{Acknowledgements}
Thanks to the fruitful discussions with Tao Xie, and Joel Larakers. 

This material is based upon work supported by the U.S. Department of Energy, Office of Science, Office of Fusion Energy Sciences, using the DIII-D National Fusion Facility, a DOE Office of Science user facility, under Award(s): 
DE-FC02-04ER54698, 
DE-SC0022164,  
DE-AC02-05CH11231, 
DE-SC0019004, 

This work was supported by U.S. DOE Contract No. DE-FG02-04ER54742 at the Instituted for Fusion Studies (IFS) at the University of Texas at Austin.

This research used resources of the National Energy Research Scientific Computing Center, a DOE Office of Science User Facility. We acknowledge the CINECA award under the ISCRA initiative, for the availability of high performance computing resources and support.

This research was supported at Oak Ridge National Laboratory supported by the Office of Science of the U.S. Department of Energy under Contract No. DE-AC05-00OR22725.

This research used resources of the National Energy Research Scientific Computing Center, a DOE Office of Science User Facility, and the Texas Advanced Computing Center (TACC) at The University of Texas at Austin.

\section{Disclaimer}
This report was prepared as an account of work sponsored by an agency of the United States Government. Neither the United States Government nor any agency thereof, nor any of their employees, makes any warranty, express or implied, or assumes any legal liability or responsibility for the accuracy, completeness, or usefulness of any information, apparatus, product, or process disclosed or represents that its use would not infringe privately owned rights. Reference herein to any specific commercial product, process, or service by trade name, trademark, manufacturer, or otherwise does not necessarily constitute or imply its endorsement, recommendation, or favoring by the United States Government or any agency thereof. The views and opinions of authors expressed herein do not necessarily state or reflect those of the United States Government or any agency thereof.

\section{Notice of copyright}

This manuscript has been authored in part by UT-Battelle, LLC, under contract DE-AC05-00OR22725 with the US Department of Energy (DOE). The publisher acknowledges the US government license to provide public access under the DOE Public Access Plan (http://energy.gov/downloads/doe-public-access-plan).

\section{Appendix}

\subsection{Simulation parameters}

For local linear simulations for 
\begin{table}[H]
    \centering
    \begin{tabular}{|c|c|c|c|c|c|c|c|}
         \hline
         & Simulation type & nx & nz & nv & nw & hypz &Shear? \\
         \hline
         & Local linear & 9 & 90 & 32 & 16 & -2 & Off \\
         \hline
         & Local nonlinear & 128 & 180 & 40 & 8 & -2 & Off \\
         \hline
\end{tabular}
\caption[font=5]{Important parameters used in the local linear and nonlinear simulations}
\label{ch:sim_para}
\end{table}

Hyperdiffusion in the z-direction is non-zero, and hyperdiffusion is zero for the rest of the directions

For ETG nonlinear simulations, ion and impurity are adiabatic, $nky=48,\ kymin=5 $. For MTM nonlinear simulation, all 3 species are non-adiabatic, $nky=24,\ kymin=0.03 $. 

A convergence test has been performed for both linear and nonlinear simulations by doubling the numerical grid points. The growth rate (for linear) and saturation level (for nonlinear) stay the same. 



Local nonlinear simulations for MTM have the following resolution:  $k_x\rho_s$ is ranging from 0 to 4 with the definition of $\frac{1}{32}$, $k_y\rho_s$ is ranging from 0 to 0.72 with the definition of 0.03 ($k_y\rho_s=n_{tor}\times0.00776$).  
The simulations have $E\times B$ shear turned off since $E\times B$ shear stabilizes the turbulence. The frequency has been translated into the lab frame during the post-processing procedure. 
Figure \ref{fig:time_trace} shows the result from local nonlinear simulation. Hyper-diffusivity in $k_x$ and $k_y$ are 0.  Hyper-diffusivity along the magnetic field is 2. After $t=55 (a/c_s)$, the low k electrostatic mode starts to run away to the nonphysical transport level. However, from the discussion in Section \ref{sec:NL_freq}, the possibility of the theory, local linear simulation, local nonlinear simulation, and experiment results tightly match at the same time by accident is little to none. Therefore, one could argue that the results from the early saturation stage of the MTM local nonlinear simulations are still valid. 


Simulating the local nonlinear MTM is a nontrivial task. The saturation level is near zero if the range of ky is too small. And one species simulation, with adiabatic ions, seems to produce a much lower transport level for MTM compared with the simulation with three species. 



\subsection{Doppler shift}
The Doppler shift must be estimated in order to make quantitative comparisons between simulation frequencies and fluctuation data.

There is no ion temperature measurement beyond $r/a\sim0.97$. The profile is fitted so that $E_r\sim0$ at the last closed flux surface (LCFS). The Doppler shift calculated for translation from plasma frame to lab frame is using the fitted profile near the edge and varies around such value within experimental uncertainty ($\pm 20\%$). 

Based on the nominal value of calculation from \cite{RIP_Chen_POP_2021}, the $v_{E\times B}\sim 10km/s$ in electron direction. Since $f_{Doppler}=\frac{v_{E\times B} B_{tor} }{2\pi R B_{pol} }$, The Doppler shift then can be calculated from the Table \ref{ch:basic_para}.  $R=1.73m$, $B_{tor}\sim1.55T$, $B_{pol}\sim0.3T$, $f_{Doppler}\sim4.75kHz$. 

On the other hand, from the neoclassical analysis, the Doppler shift for n=1 can be up to 16kHz. Therefore the Doppler shift for n=1 can range from 5kHz to 16kHz. 

\subsection{Normalization}

The collisionality is calculated as $\nu_{ei}=5 \times 10^{-11} Z_{e f f} n_{e} / T_{e}^{1.5}$ in experiment\cite{RIP_Chen_POP_2021} assuming the $ln\Lambda=17$. In GENE, the collision frequency can be expressed as $\left\langle v_{\mathrm{ei}}\right\rangle=\frac{2^{1 / 2} n_{\mathrm{i}} Z^{2} e^{4} \ln \Lambda}{12 \pi^{3 / 2} \epsilon_{0}^{2} m^{1 / 2} T_{\mathrm{e}}^{3 / 2}}$. \cite{Goldstone} The data and plots in this article have been normalized to experimental expression.

\medskip
 
\nocite{*}

\end{document}